\documentclass[aps,prx,a4paper,twocolumn,superscriptaddress,longbibliography,10pt,showkeys]{revtex4-2}
\usepackage{amsmath, braket, empheq, amsthm, amssymb, amsfonts, graphicx, color,  mathtools, geometry, collcell, placeins, mathrsfs}

\usepackage[singlelinecheck=false,caption=false]{subfig}

\usepackage{lineno}
\usepackage[table]{xcolor}
\usepackage{listings}
\usepackage{array}

\definecolor{cadmiumred}{rgb}{0.89, 0.0, 0.13}
\definecolor{intblue}{rgb}{0.0, 0.18, 0.65}

\usepackage[colorlinks=true,linkcolor=intblue,citecolor=intblue, urlcolor=black, plainpages=false,pdfpagelabels]{hyperref}

\theoremstyle{plain}
\newtheorem{thm}{\protect\theoremname}
\newtheorem{prop}[thm]{Proposition}

\providecommand{\theoremname}{Theorem}
\newcommand*{\myproofname}{Proof}
\newcommand*{\vsource}{(p(\vec{\mathcal{V}}),R_s)}
\newcommand*{\vthsource}{(p(\vec{\mathcal{V}}),R_s+R)}
\newcommand*{\Svec}{\vec{\mathcal{S}}}
\newcommand*{\rvec}{\vec{r}}

\newcommand*{\unitdc}{\mathrm{unitdc}}

\newcommand*{\unitdcdef}{\tau\left(\frac{\mathbb{E}_t (S^2) }{R_s}+4\Delta\! f \mathrm{k_B}|T_s-T|\right)}

\newenvironment{mproof}[1][\myproofname]{\begin{proof}[#1]}{\end{proof}}

\newcommand{\mc}[1]{\mathcal{#1}}
\newcommand{\msc}[1]{\mathscr{#1}}

\usepackage{tikz}
\usepackage[american]{circuitikz}
\usetikzlibrary{decorations.pathreplacing}
\definecolor{capri}{rgb}{0.0, 0.75, 1.0}
\definecolor{ceruleanblue}{rgb}{0.16, 0.32, 0.75}
\definecolor{airforceblue}{rgb}{0.36, 0.54, 0.66}
\definecolor{cinnabar}{rgb}{0.89, 0.26, 0.2}
\definecolor{columbiablue}{rgb}{0.61, 0.87, 1.0}
\definecolor{darkturquoise}{rgb}{0.0, 0.81, 0.82}
\definecolor{deepskyblue}{rgb}{0.0, 0.75, 1.0}
\definecolor{etonblue}{rgb}{0.59, 0.78, 0.64}
\definecolor{fielddrab}{rgb}{0.42, 0.33, 0.12}
\definecolor{myurlcolor}{rgb}{0,0,0.7}
\definecolor{myurlcolor1}{rgb}{0,0.7,0.1}
\definecolor{myrefcolor}{rgb}{0,0,0.7}
\usepackage{verbatim}
\usetikzlibrary{patterns}
\usetikzlibrary{graphs}
\usetikzlibrary{shadows}
\usetikzlibrary{shapes.symbols}
\definecolor{tempcolor}{RGB}{0,102,102}
\usetikzlibrary{arrows,calc}
\usetikzlibrary{shapes.arrows}
\usetikzlibrary{shapes.callouts}
\usepackage{relsize}
\usetikzlibrary{shapes}
\usetikzlibrary[arrows,decorations.pathmorphing,backgrounds,positioning,fit,petri]
\usetikzlibrary{decorations.text}

\begin{document}

\title{Quantifying the value of transient voltage sources}

\author{Swati}
\email{swati@cs.hku.hk}
\affiliation{Department of Computer Science, The University of Hong Kong, Hong Kong}
\affiliation{Shenzhen Institute for Quantum Science and Engineering and\\ Department of Physics, SUSTech, Nanshan District, Shenzhen, China}

\author{Uttam Singh}
\email{uttam@cft.edu.pl}
\affiliation{Center for Theoretical Physics PAS, Aleja Lotnik{\'o}w 32/46, 02-668 Warsaw, Poland}

\author{Oscar C. O. Dahlsten}
\email{dahlsten@sustech.edu.cn}
\affiliation{Shenzhen Institute for Quantum Science and Engineering and\\ Department of Physics, SUSTech, Nanshan District, Shenzhen, China}



\begin{abstract}
Some voltage sources are transient, lasting only for a moment of time, such as the voltage generated by converting a human motion into electricity. Such sources moreover tend to have a degree of randomness as well as internal resistance. We investigate how to put a number to how valuable a given transient source is. We derive several candidate measures via a systematic approach. We establish an inter-convertibility hierarchy between such sources, where inter-conversion means adding passive interface circuits to the sources. Resistors at the ambient temperature are at the bottom of this hierarchy and sources with low internal resistance and high internal voltages are at the top. We provide three possible measures for a given source that assign a number to the source respecting this hierarchy. One measure captures how much ``unitdc" the source contains, meaning $1$ V dc with $1\Omega$ internal resistance for $1$s.  Another measure relates to the signal-to-noise ratio of the voltage time-series whereas a third is based on the relative entropy between the voltage probability distribution and a thermal noise resistor. We argue that the unitdc measure is particularly useful by virtue of its operational interpretation in terms of the number of unit dc sources that one needs to combine to create the source or that can be distilled from the source. 
\end{abstract}



\maketitle


\section{Introduction} As the world seeks renewable and clean sources of electric energy, the importance of imperfect voltage sources increases. For example,  motion-energy-harvesting devices converting human motion to electricity are expected to play a crucial role in powering the Internet of Things without wires and batteries~\cite{Haowei2004, Mitcheson2008}. Larger-scale examples of transient sources range from regenerative braking and suspension harvesting~\cite{Abdelkareem2018, Safaei2019} to wind power~\cite{Chen2009, Blaabjerg2017}. A challenge in the usage and design of such transient voltage sources is that their voltage output often has variability, including randomness, induced by the motion source behavior, as well as significant internal resistance~\cite{halvorsen2008energy, Zhang2016, Liu2018}. 

The undeniable rise in usefulness of such transient voltage sources raises the question of whether one can capture the value of such a source in a single number, analogously to how a battery comes with a single number stating how much energy it can provide, or to the concept of free energy in thermodynamics~\cite{Feynman2018}. This is helpful both to fairly compare different source alternatives and to guide design: what number(s) should be optimized to maximize the value of a given source? One option in quantifying the value of a source is to focus on observed averages such as root-mean-squared voltage ($\mc{V}_\mathrm{rms}$)~\cite{Haowei2004, Mitcheson2008}. This does not distinguish between random or deterministic sources. Is this justified? Should one and if so by how much, penalize unpredictability as well? On one hand, thermal noise voltage should arguably be counted as having no value, but on the other hand, highly random sources with stronger noise can be used as power sources~\cite{halvorsen2008energy} and should thus carry value.  Moreover, how can one incorporate the internal resistance into the value assigned to the voltage source? 

To tackle this question of how to value transient voltage sources we make use of a conceptually clear paradigm for quantifying resources in general, known as the {\em resource theory} approach.  In our case, the resource is the voltage source.  The resource theory paradigm has proven useful to formalize thermodynamics~\cite{lieb1999physics} and is significantly used in quantum information theory~\cite{Horodecki2009, Brandao2015,  Chitambar2019}. A key idea is to define some interconversion operations as {\em free}, and consider which resources can be converted to other resources under these free operations, thus creating a hierarchy that the measure for quantifying resources must respect. If resource $A$ can be converted into resource $B$ with free operations, the number we assign to how useful $A$ is, must be at least as high as that of $B$. 

\begin{widetext}
\begin{figure*}
\centering
\label{fig:ex}
\includegraphics[width=0.8\textwidth]{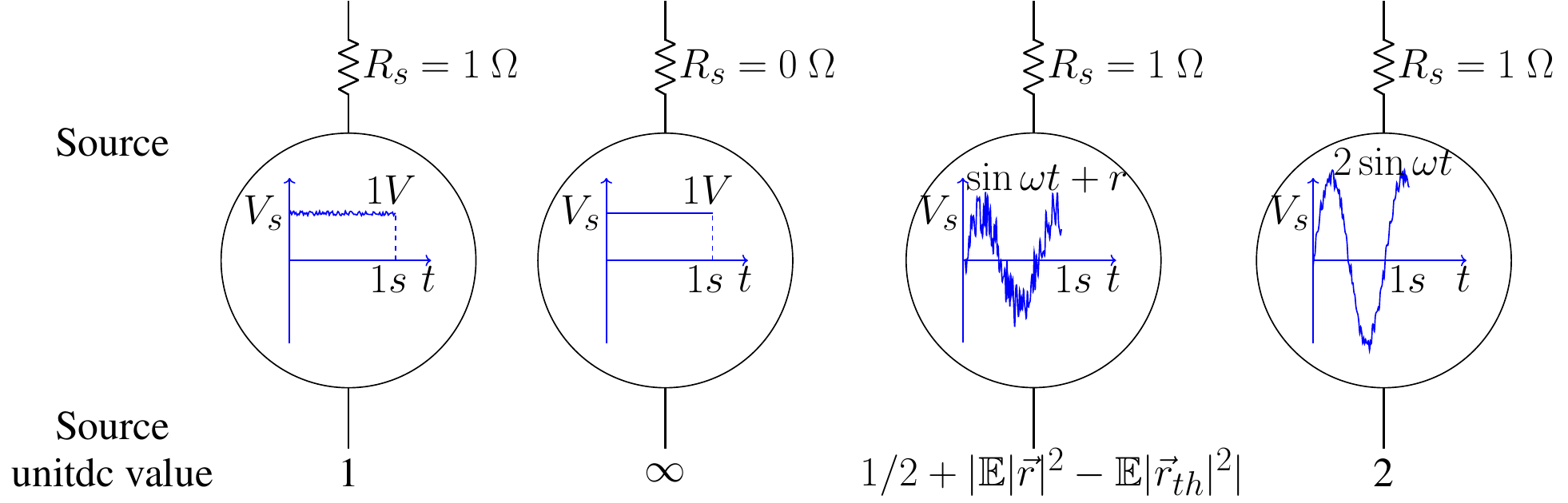}
\caption{Different transient voltage sources and their $\unitdc$ values are depicted. The sources are modeled as an internal time-dependent source voltage $V_s$ (the circle) in series with an internal resistance $R_s$ (zig-zag lines). $V_s$ is plotted as a function of time inside the circle. We assume $V_s=S+r$ where $S$ is a deterministic signal and $r$ a Gaussian random variable.  At the foot of each depicted source is the ($\unitdc$) value we assign to the given source (for a pulse duration of $1$s): $\unitdc=\frac{\tau}{R_s}\left(\mathbb{E}|S|^2 +\left|\mathbb{E}|r|^2- \mathbb{E}|r_{th}|^2\right|\right)$, where $\mathbb{E}$ denotes the statistical average of the time average over the pulse duration $\tau$, and $\mathbb{E}|r_{th}|^2$ is the variance of the noise corresponding to the internal resistance $R_s$ being at the ambient temperature (see Sec. \ref{sub:distill-dc}, Eq. \eqref{eq:val-unit}). For first and  last sources, the noise corresponds to the ambient temperature such that  $\left|\mathbb{E}|r|^2- \mathbb{E}|r_{th}|^2\right|=0$. The second source has no noise and the third has Gaussian noise that is greater than that of the ambient temperature.
We show that if a source $A$ has a higher $\unitdc$ value than another source $B$, conversion from $B$ to $A$ is not possible with `free' passive operations.}  
\end{figure*}
\end{widetext}

Here the resources $A$ and $B$ shall correspond to voltage sources and the free operations to  interconversion circuits. We treat a voltage source as a black-box two-terminal electrical device and assume it can be given a (Thevenin) equivalent circuit of a voltage source $V_s(t)$ in series with an ideal  (noncomplex) resistor. Examples of such sources are depicted in Fig. \ref{fig:ex}. 
Converting between transient voltage sources here means connecting an intermediate four-terminal circuit to the harvester, with two terminals interfacing to the harvester and another two acting as the  output. To avoid the interface circuit adding value to the original source, we want the intermediate circuit to be passive (nonpowered). More specifically, we  allow the intermediate conversion circuit to implement prechosen power-preserving reversible operations such as voltage sign flips and phase shifts~\cite{Liu2018}. At least some of those conversions are plausible to implement~\cite{Ramadass2010, Bandyopadhyay2012, Kim2013, Liang2013, Hartmann2015}. We moreover take the addition of  thermal noise at some ambient temperature $T$ as free.

Using this approach we derive three quantifiers of the value of transient voltage sources: (i) one based on the SNR of the voltage series, (ii) one which quantifies how much $1$V dc at $1\Omega$ for $1s$ one may obtain from the source, (iii) one based on the relative entropy between the source and a thermal source. We compare the three and conclude that option (ii) which we term the ``$\unitdc$" value of the source is particularly attractive due to its clear operational interpretation. Examples of sources and their $\unitdc$ values are given in  
Fig.~\ref{fig:ex}. The $\unitdc$ value moreover turns out to match well with a method used by many energy-harvesting experimentalists to quantify the performance of a given harvesting setup, in terms of the energy  dissipated into an optimal load resistance. We also show that certain alternative quantifiers, such as the peak power dissipated into a matched load, may increase under free interconversion operations and are thus arguably not good quantifiers of the value of a transient voltage source.

We proceed as follows. We begin with describing the resource theory approach concretely. Then we define the models of transient voltage sources, and the free interfacing circuits we use. We derive which resource quantifiers respect the interconversion hierarchy and which do not. We finally analyze how the value is increased under composition of sources, and how many unitdc sources can be `distilled out' from a given source via free operations.

\section{Resource theory approach}
A resource theory is a systematic framework to quantify resources at hand \cite{Chitambar2019}. Resource theories are primarily developed in the field of quantum information theory and include the resource theory of quantum entanglement~\cite{Horodecki2009}, the resource theory of quantum thermodynamics~\cite{lieb1999physics, Brandao2013}, the resource theory of asymmetry~\cite{Bartlett2007, Marvian2013, Marvian2014}, and the resource theory of coherence~\cite{Baumgratz2014, Streltsov2015}. The scope of the resource theory approach is not limited to the quantum case only as our work here exemplifies.

\begin{figure}
\centering
\includegraphics[width=65mm]{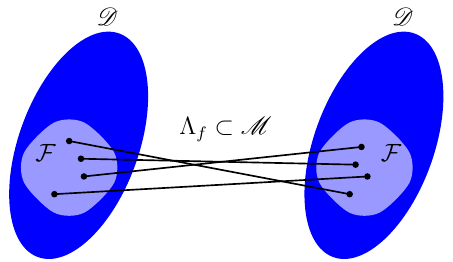}
\caption{High-level depiction of a resource theory. The set $\msc{D}$ represents the set of all states while  the set $\mathcal{F}\subset \msc{D}$ represents the set of free states. Let $\msc{M}$ denote the set of all possible operations on $\msc{D}$. Then, the free operations $\Lambda_f\subset\msc{M}$,  map $\mathcal{F}$ into $\mathcal{F}$. The remaining states $\msc{D}\setminus \mathcal{F}$ and the remaining operations $\msc{M}\setminus \Lambda_f$ are then called resource states and operations, respectively.}
\label{fig:res-theory}
\end{figure}

Let the set of all states (in our case corresponding to voltage sources) be denoted by $\msc{D}$. Let $\mathcal{V}\in \msc{D}$ be an arbitrary state. Let the set of all operations on the states be $\msc{M}$. A resource theory comprises three basic ingredients (not independent of one another):
(1) A set of free states $\mathcal{F}\subset \msc{D}$, (2) a set of free operations $\Lambda_f\subset \msc{M}$, (3) sets of resource states and operations (complementary sets to $\mathcal{F}$ and $\Lambda_f$, respectively). This is depicted in Figure \ref{fig:res-theory}.  The set of free operations is chosen such that it cannot generate resource states from the free states~\cite{Brandao2015}. 
A key purpose of setting up a resource theory is to determine a resource quantifier in a systematic manner. Any real-valued function $\mathcal{R}$ on the set of all states $\msc{D}$ is said to be a {\em resource monotone} if it satisfies the following set of conditions: \\(1) {\it Faithfulness:} $\mathcal{R} (\mathcal{V}) =0$ if and only if $\mathcal{V}$ is a free state, i.e., $\mc{V}\in\mc{F}$. \\ (2) {\it Monotonicity under free operations}:
$\mathcal{R} (\Lambda_f (\mathcal{V})) \leq \mathcal{R} (\mathcal{V})$ $\forall \, \mathcal{V}\in \msc{D}$. 

 Other conditions, e.g.\ concerning how the resource monotone behaves under probabilistic combinations of states, may also be added but are not needed here as we do not include probabilistic combinations of sources.

\section{Overview of resource theory of transient voltage sources}
\begin{itemize}
\item We identify states as voltage sources, corresponding to a pseudorandom voltage time series $V_s(t)$, placed in series with a source resistance $R_s$. Free states correspond to resistors with thermal noise at the ambient temperature. 

\item Operations are defined as connecting four-terminal circuits to the two-terminal source, such that the free two terminals are associated with the voltage output. This leads to a modified (Thevenin equivalent) source resistance $R_s'$ and a modified (Thevenin equivalent) voltage source $V_s'(t)$. Free operations correspond to certain passive and thermally noisy four-terminal circuits. 

\item We identify three resource quantifiers (monotones) respecting the interconversion hierarchy, meaning that the value the quantifier assigns to the source cannot increase under the free operations. 
\end{itemize}

Table~\ref{table:nonlin} gives an overview of the resource theory and shows the analogy with the well-known resource theory of entanglement.

\begin{widetext}
\begin{table*}[ht]
\centering
\caption{Comparison of the resource theory of transient voltage sources and the well-known resource theory of entanglement.} 
\vspace{0.5cm}
\centering 
{\setlength{\extrarowheight}{14pt}
\begin{tabular}{|p{1cm}|p{3cm}| p{6cm} | p{7cm} |} 
\hline 
\bf S.No. &\parbox[c]{3cm}{ \bf Elements}&\parbox[c]{6cm}{\bf Resource theory of entanglement} &\parbox[c]{7cm}{\bf Resource theory of transient voltage sources}\\ [2.5ex] 
\hline 
\parbox[c]{1cm}{1} & \parbox[c]{3cm}{States} & \parbox[c]{6cm}{Positive matrices with trace one \cite{Nielsen-Chuang2010}} & \parbox[c]{7cm}{($p(\vec{\mc{V}})$, $R_s$) where $p(\vec{\mc{V}})$ is a probability distribution over source voltage time series $\vec{\mc{V}}$, and $R_s$ the internal resistance}   \\[5ex] \hline
\parbox[c]{1cm}{2} &\parbox[c]{3cm}{Operations} & \parbox[c]{6cm}{Completely positive trace-preserving maps~\cite{Nielsen-Chuang2010}} & \parbox[c]{7cm}{Stochastic transformations of the voltage time series, changes to internal resistance}  \\[2.5ex] \hline
\parbox[c]{1cm}{3} &\parbox[c]{3cm}{Free states} & \parbox[c]{6cm}{Separable states \cite{Horodecki2009}} & \parbox[c]{7cm}{Gaussian white noise with zero mean at ambient temperature}  \\[2.5ex] \hline
\parbox[c]{1cm}{4 }&\parbox[c]{3cm}{Free operations} & \parbox[c]{6cm}{Local operations and classical communication (LOCC) \cite{Horodecki2009}} & \parbox[c]{7cm}{Reversible energy preserving operations and addition of thermal noise at ambient temperature}  \\[2.5ex] \hline
\parbox[c]{1cm}{5} &\parbox[c]{3cm}{Resource states} & \parbox[c]{6cm}{Entangled states} & \parbox[c]{7cm}{Voltage sources that are not thermal noise sources at the ambient temperature}  \\ [2.5ex] \hline
\parbox[c]{1cm}{6} &\parbox[c]{3cm}{Composition} & \parbox[c]{6cm}{Tensor product} &\parbox[c]{6cm}{$\mathcal{V}+\mathcal{V}'$ (sources in series)}\\[2.5ex] \hline
\parbox[c]{1cm}{7} &\parbox[c]{3cm}{Typical resource monotones} & \parbox[c]{6cm}{Relative entropy of entanglement, distillable entanglement, and entanglement cost }&\parbox[c]{7cm}{ $\mathrm{MSNR}$, relative entropy and $\unitdc$}\\[5ex] 
\hline 
\end{tabular}}
\label{table:nonlin} 
\end{table*}
\end{widetext}

We now proceed to describe this resource theory concretely, beginning with the states.

\section{Model for transient voltage source: The states}

We model a two-terminal voltage source as an ideal voltage source in series with a resistor. An ideal voltage source alone yields, by definition~\cite{horowitz1989art}, a voltage across its two terminals that is independent of what they are connected to. This is an unphysical idealization and real devices, such as real batteries~\cite{panasonic}, are often modeled as a voltage source with voltage $V_s$ in series with a resistor with resistance $R_s$. A justification for this model is that physical circuits are known to be well represented by networks of resistors, capacitors, and inductors (the lumped-element model) and within that model, it can be shown that a two-terminal circuit acts, at the black-box level, identically to an ideal voltage source in series with an (in general complex) impedance $Z_s$~\cite{horowitz1989art}. Here we choose for technical convenience $Z_s$ to be real~\footnote{If $Z_s$ is complex, the source will arguably have the same value as a source with $R_s=\mathrm{Re}(Z_s)$, at least in the case of a sinusoidal single-frequency voltage $V_s$. In that case, and possibly in general, one  can (for idealised circuits) reversibly convert between $R_s=\mathrm{Re}(Z_s)$ and the complex impedance $Z_s$ using an LC circuit (which is passive) connected in series. (An inductor and capacitor in series have $Z_{LC}(\omega)=i(\omega L)-i\frac{1}{\omega C}$ which can thus be made to equal $-\mathrm{Im}(Z_s)$ by suitable choice of constants $L$ and $C$.)}.

The source voltage $V_s$ and resistance $R_s$ can be determined experimentally. If connected to a resistive load of resistance $R_L$ then the voltage across the two terminals equals (up to a sign) that across the load, namely $V_L=IR_L=\frac{V_s}{(R_s+R_L)}R_L=\frac{V_s}{(\frac{R_s}{R_L}+1)}$, where $I=\frac{V_s}{(R_s+R_L)}$ is the current through the load. One sees $I$ is impeded by the internal resistance $R_s$. That expression for $V_L$ moreover justifies a method for experimentally determining $R_s$ and $V_s$: checking the voltage drop when connecting a load. The open-circuit measured voltage $V_{oc}$ is approximately the load voltage associated with $R_L\rightarrow \infty$ so $V_{oc}=V_s$. Moreover, in a few lines, one sees $R_s=\frac{V_s}{I}-R_L=\frac{V_s}{I}-\frac{V_L}{I}$ for which all the terms in the rhs are experimentally accessible with a voltmeter and ammeter. This procedure is analyzed in the context of real batteries, e.g.\ in~\cite{panasonic}. $V_s$ moreover has an important operational meaning in terms of the black-box voltage $V_L$ for an optimized resistive load with resistance $R_L$.  It is in fact common in experiments in energy harvesting~\cite{Mitcheson2008} to compare different harvesters by the energy dissipated in an optimized resistive load, with $R_L$ chosen such that it optimizes the power dissipated in the load: $P_L=V_L^2/R_L$. One can see that for all other parameters being equal, $P_L$ is optimized for $R_L=R_s$, by noting that $V_L=IR_L=\frac{V_sR_L}{(R_s+R_L)}$ and setting $\partial_{R_L}P_L=0$. Thus $V_L$ evaluated at $R_L=R_S$ obeys
\begin{equation}
\label{eq:vlvs}
V_L|_{R_L=R_s}=\mathrm{argmax}_{V_L}P_L=\frac{V_s}{2}.
\end{equation} 
It follows that, up to a factor of 2, the internal source voltage $V_s$ may be thought of as the load voltage $V_L$ for an optimized resistive load. Finally, we note that, in practice, $V_s$ will be sampled some $N$ times at fixed time intervals $\delta t$, yielding a vector $\vec{\mc{V}}:=(V_1,\cdots, V_N)^{\dagger}$, where $\dagger$ indicates the transpose.

We allow for $\vec{\mc{V}}$ to be probabilistic, such that a source is characterized by two things (i) the probability distribution $p(\vec{\mc{V}})$ and (ii) the internal resistance $R_s$. In other words, a source is a 2-tuple ($p(\vec{\mc{V}})$, $R_s$). 

We now describe the model for $p(\vec{\mc{V}})$. The source voltage $V_s(t)$ may depend on time and last for some specified time interval.  Secondly, $V_s(t)$ may be {\em noisy}, a subtle notion. From the perspective of a given agent, a time series may be split, in line with the signal-processing paradigm, into {\em signal} plus {\em noise}: $V_i=S_i+r_i$, where $S_i$ is deterministic and $r_i$ random and $i$ labels the sample. For example we may have a sinusoidal signal such that $S_i=\sin(\omega t_i)$ and $r_i$ being picked randomly from a normal distribution (Gaussian) at each $i$. We allow for the signal to be a Fourier series in general. For this class of generating models there are thus well-defined vectors $\vec{\mc{S}}:=(S_1,\cdots, S_N)^{\dagger}$ and $\vec{r}:=(r_1,\cdots, r_N)^{\dagger}$ for any run of an experiment. Moreover the distribution of $\vec{\mc{V}}$ is determined by those of $\vec{\mc{S}}$ and $\vec{r}$. In the example when the signal is sinusoidal and the noise is Gaussian, 
$\vec{\mathcal{S}}$ is distributed as
\begin{align}
\label{eq:prob-s}
p(\vec{\mathcal{S}})= \prod_i p_i(\mathcal{S}_i) =\prod_i  \delta(\mathcal{S}_i - \sin\omega t_i ),
\end{align}
where $\delta(\cdot)$ is the Dirac $\delta$ function and the noise $\vec{r}$ as
\begin{align}
\label{eq:prob-r}
q(\vec{r}) = \prod_i q(r_i)=\prod_i  \frac{1}{\sqrt{2\pi \sigma_i^2}} \exp\left[-\frac{r_i^2}{2\sigma_i^2}\right].
\end{align}

\begin{figure}
\centering
\includegraphics[width=65 mm]{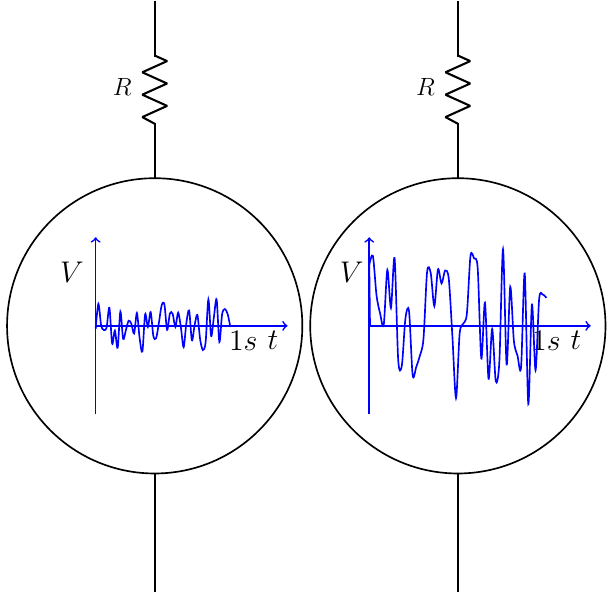}
\caption{Two ideal resistors with Gaussian voltage time series. Each models a single physical resistor at some temperature. At each time moment, a voltage value is selected from a Gaussian distribution centred on 0 and whose standard deviation is proportional (for some frequency window) to $RT$ where $R$ is the resistance and $T$ the temperature of the thermal resistor. Thus the right resistor has a higher temperature than the left. }
\label{fig:gaussian}
\end{figure}

Such noise is termed additive white Gaussian noise (`additive' because $V_i=S_i+r_i$, `white' because the power spectral density of $r(t)$ turns out to be the same for all frequencies), abbreviated as AWGN. This choice of a Gaussian noise model is common, convenient, and in our case particularly useful because we want the resource hierarchy we establish to be consistent with thermodynamics, such that thermal resistors at the ambient temperature $T$ are at the bottom of the hierarchy. Whilst for general sources $\vec{\mc{V}}$ may be independent of the internal resistance $R_s$, the voltage generated by thermal resistors, a special type of source, is commonly modeled as being uniquely determined by their resistance 
and the ambient temperature $T$. A physical resistor at temperature $T$ is then modeled mathematically as an ideal resistor in series with a voltage source $V_T$ sampled as additive white Gaussian noise (AWGN) with variance given by~\cite{Johnson1928, Nyquist1928, Schmid1982, Kish2002, Zon2004, Garnier2005, Niskanen2007, Burkard2004, Clerk2010, Bergeal2010, Seifert2012, Ciliberto2013, Ciliberto2013b, Delvenne2014, Pekola2015, Macklin2015, Solgun2015, Rao2018, Rodriguez2019, Freitas2020, Freitas2021}
\begin{equation}
\label{eq:Johnson-Nyquist}
\sigma^2=\Delta f 4 k_B TR_T,
\end{equation}
where $\Delta f$ is the frequency bandwidth being sampled, which can be thought of as a setting in an oscilloscope and $k_B$ is Boltzmann's constant. The subscript $T$ in $R_T$ and $V_T$ indicates that this is the resistance of a thermal resistor and the voltage across that resistor respectively. [Strictly speaking, Eq.~\eqref{eq:Johnson-Nyquist} is an approximation that leads to a variance that diverges in $\Delta f$ and more generally the Planck distribution should be used~\cite{Johnson1928}.] Such thermal resistors are depicted in Fig.~\ref{fig:gaussian}. There may moreover be Gaussian noise that comes not from a thermal process, but e.g.\ due to a variety of factors giving rise to Gaussian noise via the (Lindeberg) central limit theorem for independent random variables~\footnote{For the Lindeberg central limit theorem see e.g.\ \url{https://www.stat.berkeley.edu/users/pitman/s205f02/lecture10.pdf}}. In this case, we may interpret the variance of the source voltage as giving rise to an {\em effective} temperature of the noise. For many practical energy-harvesting applications, the genuinely thermal noise is indeed insignificant. A $10k\Omega$ resistor at room temperature has an open-circuit rms voltage of $1.3\mu$V, measured with a bandwidth of $10$ kHz~\cite{horowitz1989art} whereas the output of hand-sized harvesters is often of the order of volts or higher. Thus for many applications of the results here we expect any noteworthy Gaussian noise not to be thermal and the variance thus to be associated with an {\em effective} temperature.

If the generating model is {\em a priori} not known, one may rather repeat the experiment many, say $n$ times to build up a set  $\{\vec{\mc{V}}_k\}_{k=1}^n$, assume it is generated by a distribution within a family of models, and match the data to the `best' model within that family via the maximum-likelihood method, as described in~Appendix \ref{sec:model}. 


We take any resistor with resistance $R$ at ambient temperature $T$ to be a {\em free state}. All other voltage sources, including resistors at other temperatures, are taken as nonfree states. 

\section{Inter-conversion Operations}
We now describe the free interconversion operations between sources that we allow.

\subsection{Reversible energy-preserving operations} 
Motivated by basic physics and previous work~\cite{Liu2018} we define one type of free operations as being ones which are reversible and energy preserving, (without changing the internal resistance $R_s$). Here {\em reversible} means that the inverse operation is also allowed (and free). By {\em energy preserving} we mean that the energy delivered to a resistive load,  $\frac{\vec{\mc{V}_L}\cdot \vec{\mc{V}_L}}{NR_L}$, is conserved, where $N$ is the number of sampled points. For fixed $R_L$ and $R_s$, $V_L$ and $V_s:=V$ are proportional (recall $V_L=\frac{V_s}{1+\frac{R_s}{R_L}}$). Thus, the demand that 
$\vec{\mc{V}_L}\cdot \vec{\mc{V}_L}$ is conserved under these operations is equivalent to demanding that $\vec{\mc{V}}\cdot \vec{\mc{V}}$ is conserved.

Matrices, which preserve dot products, are orthogonal. Modeling the operations as matrices implies that if sources are added in series, the impact of the operation is the same as if it had acted on each source individually. Moreover, multiplication by a factor commutes with applying an orthogonal matrix, so we may treat the matrix as acting on either $\vec{\mc{V}}$ or $\vec{\mc{V}_L}$ and choose the former, since we opt to represent the source as the 2-tuple ($p(\vec{\mc{V}})$, $R_s$). We thus model these energy-preserving reversible operations as orthogonal matrices $O$ acting on $\vec{\mc{V}}$. 

There is no measurement during the implementation of the free operations. There may have been measurements involved in characterizing the source's probability distribution (a process sometimes called  tomography) but we do not include the one-off cost of those.

\begin{figure}
\centering
\includegraphics[width=65mm]{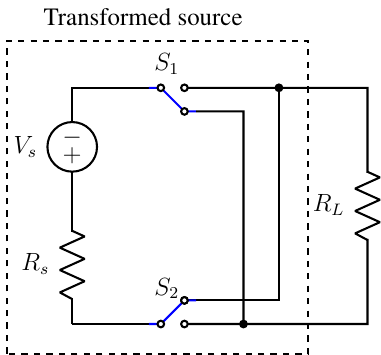}
\caption{Example of converting one source to another. There is a nonideal random voltage source with associated resistance $R_S$. There is a load, often taken to be purely resistive with resistance $R_L$. The source may be transformed by certain (in principle) energy-preserving reversible operations such as flipping $V\mapsto -V$ by switching the circuit architecture facilitated by switches $S_1$ and $S_2$ as depicted.}
\label{fig:biasflip}
\end{figure}

A simple example of such operations is the (possibly periodic) bias flip~\cite{Li2016, Liu2018}. This is depicted in Fig.~\ref{fig:biasflip}. If the aim of the agent is, e.g.\ to outperform the diode bridge for the task of rectifying the voltage with minimal loss, the bias flip can, in principle, succeed provided that the signal-to-noise ratio is not too high~\cite{Liu2018}. (Rectification is further discussed in~Appendix \ref{append:sec-dist}.) Another simple example is a phase shift, which shifts the time label along. As toy examples of these, consider a voltage time series $\vec{\mathcal{V}}=(V_1,V_2,V_3,V_4)^{\dagger}$. An example of the bias-flip transformation denoted by $O_{BF}$ and a phase-shift transformation denoted by $O_{PS}$ is given, respectively, by the following $4\times 4$ orthogonal matrices:
\begin{align*}
O_{BF} = 
\begin{pmatrix}
1&0 &0&0\\
0&-1 &0 &0\\
0&0 &1 &0\\
0&0 &0 &-1
\end{pmatrix} ~\mathrm{and}~ O_{PS} = 
\begin{pmatrix}
0&1 &0&0\\
0&0 &1 &0\\
0&0 &0 &1\\
1&0 &0 &0
\end{pmatrix}.
\end{align*}

More generally, the voltage time-series transform under the energy-preserving reversible operations as
\begin{align}
\label{volt2}
\tilde{V}_i=\sum_j O_{ij}V_{j} =  \sum_j O_{ij} \mathcal{S}_j + \sum_j O_{ij} r_j,
\end{align}
where $O$ is an orthogonal matrix.

\subsection{Adding thermal noise} 
\begin{figure}[htbp!]
\begin{center}
\includegraphics[width=85mm]{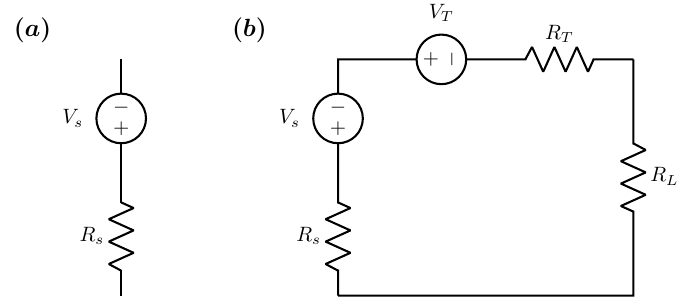}
\caption{The free operation of adding a thermal source. Subfigure (a) represents an open-circuit nonideal voltage source modeled as an ideal voltage source $V_s$ connected in series with a resistance $R_s$. Subfigure (b) represents a closed circuit nonideal voltage source,  which has thermal noise added to it, thus becoming an alternative  transformed source. The nonideal voltage source is modeled as an ideal voltage source $V_s$ connected in series with a resistance $R_s$ and the thermal noise is modeled as an ideal voltage source $V_T$ connected in series with a resistance $R_T$. Moreover, from the Johnson-Nyquist formula [Eq. \eqref{eq:Johnson-Nyquist}, Refs.~\cite{Johnson1928, Nyquist1928}], we have $\mathbb{E} |V_T|^2 \propto R_T T$, where $T$ is the ambient temperature for the thermal noise. Subfigure (b) also depicts a load resistance $R_L$.}
\label{fig-non-ideal-model}
\end{center}
\end{figure}

Adding a thermal resistor $R_T$ at the ambient temperature $T$ will also be treated as a free operation. The variance of such a resistor's $V_s$ is given by Eq.~\eqref{eq:Johnson-Nyquist} and is proportional to $R_T T$. Thus the addition is modeled as
$\vsource \mapsto (p'(\vec{\mc{V}}), R_s+R_T)$, where $p'$ represents a process where AWGN noise satisfying Eq.~\eqref{eq:Johnson-Nyquist} has been added. This free operation is depicted in Fig.~\ref{fig-non-ideal-model}.

\section{Which resource quantifiers respect conversion hierarchy}
We now use the resource theory paradigm to identify quantities that respect the hierarchy associated with the free operations. If one source can be converted to another with free operations the latter source must not have a higher number assigned to it. Quantities respecting the conversion hierarchy in that sense are then reasonable quantifiers of the value of transient voltage sources. 

The value given by the quantifiers involves taking averages. To explain the notation we use, for example, a vector $\vec{u}=\left(u_1,\cdots,u_N\right)^{\dagger}$. Assume it is sampled at $N$ intervals of $\delta t$ over a total time $\tau=N\delta t$. It is natural to define the time average $\mathbb{E}_t(u)=\frac{\sum_i u_i\delta t}{\tau}=\frac{\sum_i u_i}{N}$. Then, for example,  $\mathbb{E}_t(u^2)=\frac{\vec{u}\cdot \vec{u}}{N}$. We moreover employ the statistical average $\mathbb{E}_p(u)=\sum_u p(u)u$. The statistical average of a time average will also appear, e.g.\ $\mathbb{E}_{p,t}(u^2)=\mathbb{E}_p(\frac{\vec{u}\cdot \vec{u}}{N}):=\mathbb{E}|\vec{u}|^2$.

We begin with discussing quantifiers that fail to respect the hierarchy before giving three that respect it.

\begin{figure}
\centering
\includegraphics[width=85 mm]{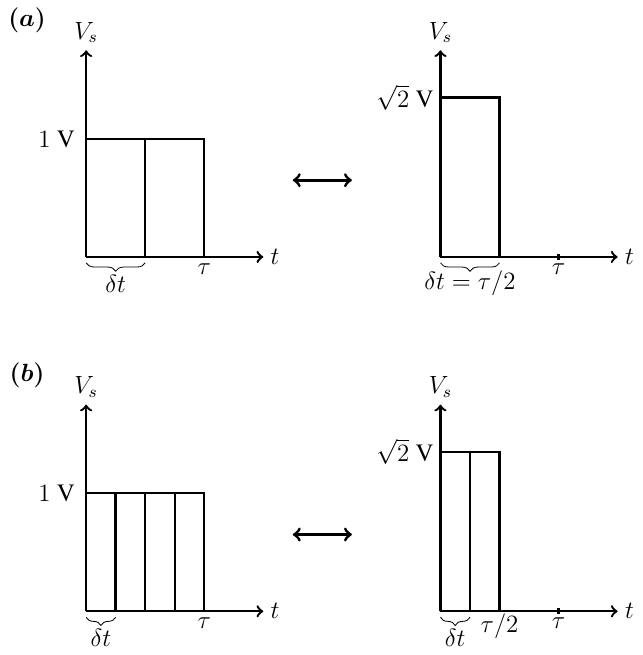}
\caption{Toy example of a free conversion that changes the time-maximized power. The left parts of both subfigures ((a) and (b)) depict $1$ V deterministic voltage source for a duration of $\tau=1$s. The right parts of both subfigures ((a) and (b)) depict $\sqrt{2}$V deterministic voltage source for a duration of $\tau=1$s.(a) In this case, the source is sampled into intervals of $\delta t=1/2$s. Thus, the voltage vectors, in this case, for left and right subfigures are given by $2\times 1$ vectors $\vec{u}=(1, 1)^{\dagger}$ and $\vec{v}=\sqrt{2}(1, 0)^{\dagger}$, respectively. The Hadamard matrix $H$, which is an orthogonal matrix, is then used to convert $\vec{u}$ to $\vec{v}$, i.e.,  $\vec{v}=H\vec{u}$ (see Appendix \ref{append:ex-orth-matrix}). (b) In this case, the source is sampled into intervals of $\delta t=1/4$s. Thus, the voltage vectors, in this case, for left and right subfigures are given by $4\times 1$ vectors $\vec{u}'=(1, 1, 1, 1)^{\dagger}$ and $\vec{v}'=\sqrt{2}(1, 1, 0, 0)^{\dagger}$, respectively. The $4\times 4$ orthogonal transformation $O$ such that $\vec{v}'=O\vec{u}'$ is provided in Eq. \eqref{eq:4d-case-orth} of Appendix \ref{append:ex-orth-matrix}.}.\label{fig:ex-orth}
\end{figure}

\subsection{Maximal power dissipated does not respect the hierarchy}
A common measure in the literature on energy harvesting turns out not to be a monotone under the free operations of our resource theory. Suppose the experimenter tunes the resistive load $R_L$ until maximum power is dissipated and then records the power for the optimal resistance and also, crucially, maximizes it over time, i.e.\ $\max_{R_L, t}P_L$, where $P_L=\frac{V_L^2}{R_L}$. From Eq.~\eqref{eq:vlvs} one then sees $\max_{R_L, t}P_L=\max_t\frac{{V_s^2(t)}}{4R_s}$. This quantity is not monotonic under our free operations, however. For example, if $V_s=1V$ for 1 second, then we allow for the free interconversion via orthogonal operations to $V_s=\sqrt{2}V$ for half a second, which would increase
$\max_{R_L, t}P_L$. The construction of the orthogonal matrix achieving this interconversion is presented in Appendix \ref{append:ex-orth-matrix} (also see Fig. \ref{fig:ex-orth}).

Moreover, recall that we allow adding thermal noise as a free operation. This could increase $\max_{R_L, t}P_L$ as well, e.g.\ if initially $V_s=0$. This addition would even increase the energy delivered to a resistive load, $\mathbb{E}_p P_L\tau$. Thus the energy delivered to a resistive load is not a monotone either. Similarly $\max_{R_L}\mathbb{E}_p P_L\tau$, also used in the energy-harvesting literature is strictly speaking not a monotone, except for, notably, being an approximate monotone in the regime where thermal noise is insignificant relative to the other voltages. 

The above argument formalizes why we should deduct the ambient temperature thermal noise from the signal before evaluating the usefulness of the transient voltage source: adding ambient temperature thermal noise should not increase the source value. In what follows we give three quantities that involve this deduction and are indeed resource monotones (nonincreasing under free operations).

\subsection{Modified signal-to-noise ratio}
\label{sec:snr}
The standard SNR is defined as $\frac{\mathbb{E}|\vec{S}|^2}{\mathbb{E}|\vec{r}|^2}$ for signal vector $\vec{S}$ and noise vector $\vec{r}$. SNR is widely used in information processing and information theory so it is arguably natural to consider whether a voltage source resource quantifier can be built from it. 

We firstly propose a modified SNR $(\mathrm{MSNR})$, as a possible resource monotone. Let 
\begin{align}
\label{MRSNR}
\mathrm{MSNR}\vsource:=\frac{\mathbb{E}|\vec{S}|^2+\left|\mathbb{E}|\vec{r}|^2-\mathbb{E}|\vec{r}_{th}|^2\right|}{R_s\mathbb{E}|\vec{r}|^2}\tau,
\end{align}
where $ \vec{r}_{th}$ is the thermal noise corresponding to $R_s$ at ambient temperature, i.e.\ $\mathbb{E}|\vec{r}_{th}|^2=\Delta f 4 k_B TR_s$ using Eq.~\eqref{eq:Johnson-Nyquist},  and the averages ($\mathbb{E}$) are taken as statistical averages of time averages over the probability distributions of the corresponding signal and noise sources as defined in the beginning of the section. $\tau$ is the pulse duration of the source, which is natural to include since a longer pulse with a given average source voltage should be expected to be more valuable than a short one.  The reason to include the internal resistance $R_s$ is to ensure that our monotone adds extra penalty on the sources that have higher internal resistance. The  $ \vec{r}_{th}$ term, associated with the ambient temperature, ensures that resistors that are hotter and colder than the ambient temperature resistors, remain nontrivial sources.  MSNR, given by Eq.~\eqref{MRSNR}, does satisfy the two requirements we list for a resource monotone.

\begin{itemize}
    \item {\it{Faithfulness:}} It immediately follows from the definition that $\mathrm{MSNR}$ is zero for the free states ($\Svec=0, ~\vec{r}=\vec{r}_{th}$) and vice versa, and is thus a faithful quantifier. 
    Resistors at a different temperature to the ambient one have $\mathrm{MSNR}\neq 0$ since, although $\Svec=0$, $\vec{r}\neq \vec{r}_{th}$.
    
    \item {\it{Monotonicity under free operations:}}  MSNR is conserved, and thus nonincreasing, under orthogonal transformations for which $\tilde{\mathcal{S}}_i= \sum_j O_{ij} \mathcal{S}_j$ and $\tilde{r}_i = \sum_j O_{ij} r_j$ where $O$ is an orthogonal matrix. This can be shown as follows:
\begin{align}
\label{eq:inv-ort1}
\sum_i \tilde{\mathcal{S}}_i^2 &= \sum_i \sum_j O_{ij} \mathcal{S}_j \sum_k O_{ik} \mathcal{S}_k\nonumber\\
&=  \sum_{j,k} \mathcal{S}_j \mathcal{S}_k \delta_{j,k} =  \sum_{j}  \mathcal{S}_j^2.
\end{align}
By the same argument, under orthogonal transformations  $\sum_i \tilde{r}_i^2=\sum_{j}  r_j^2$ and $\sum_i \tilde{r}_{th,i}^2=\sum_{j}  r_{th,j}^2$.
Moreover, since adding thermal noise at the ambient temperature, the other free operation, increases $\mathbb{E}(|\vec{r}|^2)$ whilst not impacting the signal, it is clear that $\mathrm{MSNR}$ will not increase under such operations. Under addition of thermal noise at ambient temperature it is easy to see that the monotone is nonincreasing. When adding a thermal resistor $R$ in series $\vec{r}\mapsto \vec{r}+\vec{s}_{th},~\vec{r}_{th}\mapsto \vec{r}_{th}+\vec{s}_{th}~\text{and}~R_s\mapsto R_s+R$, where $\vec{s}_{th}$ is the thermal noise at ambient temperature corresponding to resistor with resistance $R$. Then the MSNR value of Eq.~\eqref{MRSNR} becomes 
\begin{align}
\label{mon-MRSNR}
&\frac{\mathbb{E}|\Svec|^2+\left|\mathbb{E}|\vec{r}+\vec{s}_{th}|^2-\mathbb{E}|\vec{r}_{th}+\vec{s}_{th}\right|^2|}{(R_s+R)~\mathbb{E}|\vec{r}+\vec{s}_{th}|^2} \tau\nonumber\\
&=   \frac{\mathbb{E}|\Svec|^2+\left|\mathbb{E}|\vec{r}|^2-\mathbb{E}|\vec{r}_{th}|^2\right|}{(R_s+R)~\mathbb{E}|\vec{r}+\vec{s}_{th}|^2}\tau \nonumber\\
& \leq \frac{\mathbb{E}|\Svec|^2+\left|\mathbb{E}|\vec{r}|^2-\mathbb{E}|\vec{r}_{th}|^2\right|}{R_s~\mathbb{E}|\vec{r}|^2}\tau.
\end{align}
From Eq.~\eqref{eq:inv-ort1} and Eq.~\eqref{mon-MRSNR} we see that the resource value assigned by MSNR is nonincreasing under the action of free operations and is hence a valid monotone.
\end{itemize}
      
MSNR may also be conveniently evaluated in several cases. Table \ref{table:RSNR} gives the MSNR for different sources $\vsource$, with $R_s$ kept general with AWGN noise (Eq. \eqref{eq:prob-r}) that has time-dependent variance.
\FloatBarrier
\begin{table}[ht]
 \caption{MSNR for different sources $\vsource$. It is assumed that signal dominates the noise term, $\mathbb{E}|\Svec|^2\gg \max(\mathbb{E}|\vec{r}|^2,\mathbb{E}|\vec{r}_{th}|^2)$ such that only the noise in the denominator needs to be included. The AWGN noise standard deviation is allowed to depend on time, labeled as $\sigma_i$. The signal duration $\tau=1$. The corresponding proofs are given in~Appendix \ref{append-proofs}. Derivations of the MSNR value for broadband sources are also given in ~Appendix \ref{append-snr-thm-broadband}.} 

\vspace{0.5cm}
\centering 
 \begin{tabular}{|p{0.1\textwidth}|p{0.38\textwidth}| } 
 \hline 
 {\bf Quantity} & {\bf Value}\\
 
 \hline 
 \hline
 $\Svec_i$ & $\sin(\omega t_i)$\\
 \hline
 $\rvec$   & Thermal AWGN, $\sigma_k$ at time $t_k$\\
 \hline
 MSNR  &$\left(2R_s \lim_{N\rightarrow\infty}\frac{1}{N} \sum_{k=1}^N \sigma_k^2\right)^{-1}$~~~(Eq. \eqref{eq:simp-form})\\
 \hline
 \hline 
 $\Svec_i$ & $\sin(\omega t_i) + \cos(\omega t_i)$\\
 \hline
 $\rvec$   & Thermal AWGN, $\sigma_k$ at time $t_k$ \\
 \hline
 MSNR  & $\left(R_s \lim_{N\rightarrow\infty}\frac{1}{N} \sum_{k=1}^N \sigma_k^2\right)^{-1}$~~~(Eq. \eqref{eq:sin-cos})\\
 \hline
 \hline 
 $\Svec_i$ & $\sum_{\omega'} \left(f(\omega')\sin\omega' t_i + g(\omega')\cos\omega' t_i\right)$\\
 \hline
 $\rvec$   & Thermal AWGN, $\sigma_k$ at time $t_k$ \\
 \hline
 MSNR  & $\left(\sum_{k=1}^N \left[\sum_{\omega'} \left(f(\omega')\sin\omega' t_k \!+\! g(\omega')\cos\omega' t_k\right)\right]^2\right)\times$ $\left(R_s\sum_{k=1}^N \sigma_k^2\right)^{-1}$~~~(Eq. \eqref{Eqn:avgbroad})\\
\hline
\end{tabular}
\label{table:RSNR} 
\end{table}

 We show that MSNR assigns values to sources in a way that respects the hierarchy, and that it can be convenient to calculate. However, when there is no noise ($r=\sigma \approx 0$), MSNR diverges even for finite $R_s$, suggesting the MSNR value itself may not have a direct operational meaning, such as the number of unit resources that can be distilled. We therefore propose and investigate a second resource monotone, which turns out to have such an operational meaning.\\

\subsection{\texorpdfstring{$\unitdc$}{}}
\label{sub:distill-dc}
There exists a resource monotone whose value, we show, can be interpreted as the number of unit dc sources (1V at 1$\Omega$ for 1s) contained within the source. We accordingly denote this quantifier by $\unitdc$. 
We let
\begin{align}
\label{eq:val-unit}
\unitdc\vsource :=
\frac{\mathbb{E}| \Svec |^2 +\left| \mathbb{E}| \vec{r}|^2 - \mathbb{E}| \vec{r}_{th}|^2\right|}{ R_s} \tau,
\end{align}
where $\Svec$ is the signal, $ \vec{r}$ is the noise part in the signal, $R_s$ the internal resistance of the source, $ \vec{r}_{th}$ is the thermal noise corresponding to $R_s$ at ambient temperature and the averages ($\mathbb{E}$) are taken as statistical averages of time averages over the probability distributions of the corresponding signal and noise sources as defined in the beginning of the section. $\tau$ is the pulse duration of the  source. The physical unit of $\unitdc$ is energy (J). It can be seen easily that $\unitdc$ is a monotone.

\begin{itemize}
\item {\it Faithfulness:} By inspection $\unitdc\vsource=0$ if and only if $\vec{\mathcal{V}}$ has no signal and noise is only thermal at ambient temperature: $\Svec=\vec{0}$ and $\vec{r} = \vec{r}_{th}$. This is true for our free states. It is also clear that for a hotter or cooler than ambient temperature resistor, even though $\Svec=0$, the $\unitdc\neq 0$ as $| \vec{r}|^2 \neq | \vec{r}_{th}|^2$. 

\item {\it Monotonicity under free operations:} $\unitdc\vsource$ is conserved, and thus nonincreasing, under orthogonal transformations since they preserve  $|\Svec|^2$, $|\vec{r}|^2$, and $|\vec{r}_{th}|^2$ (see Eqs. \eqref{eq:inv-ort1}). The other free operation is to add thermal noise at the ambient temperature, which is equivalent to adding a resistor $R$ in series  at the ambient temperature. Thus, the modified internal resistance becomes $R_s'=R_s+R>R_s$. Such a resistor adds an independently distributed Gaussian noise $\vec{s}_{th}$ whose variance is proportional to $Rk_B T$.  Thus, we have
\begin{align}
&\unitdc\vthsource\nonumber\\
&=
\frac{\mathbb{E}| \Svec |^2 +\left| \mathbb{E}| \vec{r}+\vec{s}_{th}|^2 - \mathbb{E}| \vec{r}_{th} + \vec{s}_{th}|^2\right|}{(R_s+R)} \tau\nonumber\\
&=
\frac{\mathbb{E}| \Svec |^2 +\left|\mathbb{E} | \vec{r}|^2 - \mathbb{E}| \vec{r}_{th} |^2\right|}{(R_s+R)} \tau\nonumber\\
&\leq \unitdc\vsource.
\end{align}
Thus, adding thermal noise at the ambient temperature does not increase $\unitdc\vsource$ and we conclude that  $\unitdc\vsource$ is indeed a monotone under free operations.
\end{itemize}

\noindent
The value of $\unitdc\vsource$ can be evaluated using similar techniques described above for MSNR. 



A motivation for using $\unitdc\vsource$ over other monotones is that it can be interpreted as the amount of `unit dc' sources within the source. These are 1 V dc sources with internal resistance $R_s=1\Omega$, pulse duration $\tau=1s$, thermal noise corresponding to $1\Omega$ resistance at the ambient temperature, and are represented as an $N$ length time series
\begin{align}
\label{eq:val-one-dc1}
\vec{\mathcal{V}}_{dc} = \vec{e} + \vec{r}_{th},
\end{align}
where $\vec{e}=(\overbrace{1,\cdots,1}^N)^{\dagger}$ such that the probability distribution $p(\vec{e})=\prod_i \delta(e_i-1)$. For such a source,
\begin{align}
\label{eq:val-one-dc}
\unitdc(p(\vec{\mathcal{V}}_{dc}),1)&=\mathbb{E}| \vec{e} |^2\nonumber\\
&=\frac{1}{N} \sum_{e_1,\cdots,e_n} \left( \prod_{j=1}^N\delta(e_j-1) \right)\sum_{i=1}^N e_i^2\nonumber\\
&=1.
\end{align}

\begin{figure}
\centering
\includegraphics[width= 85 mm]{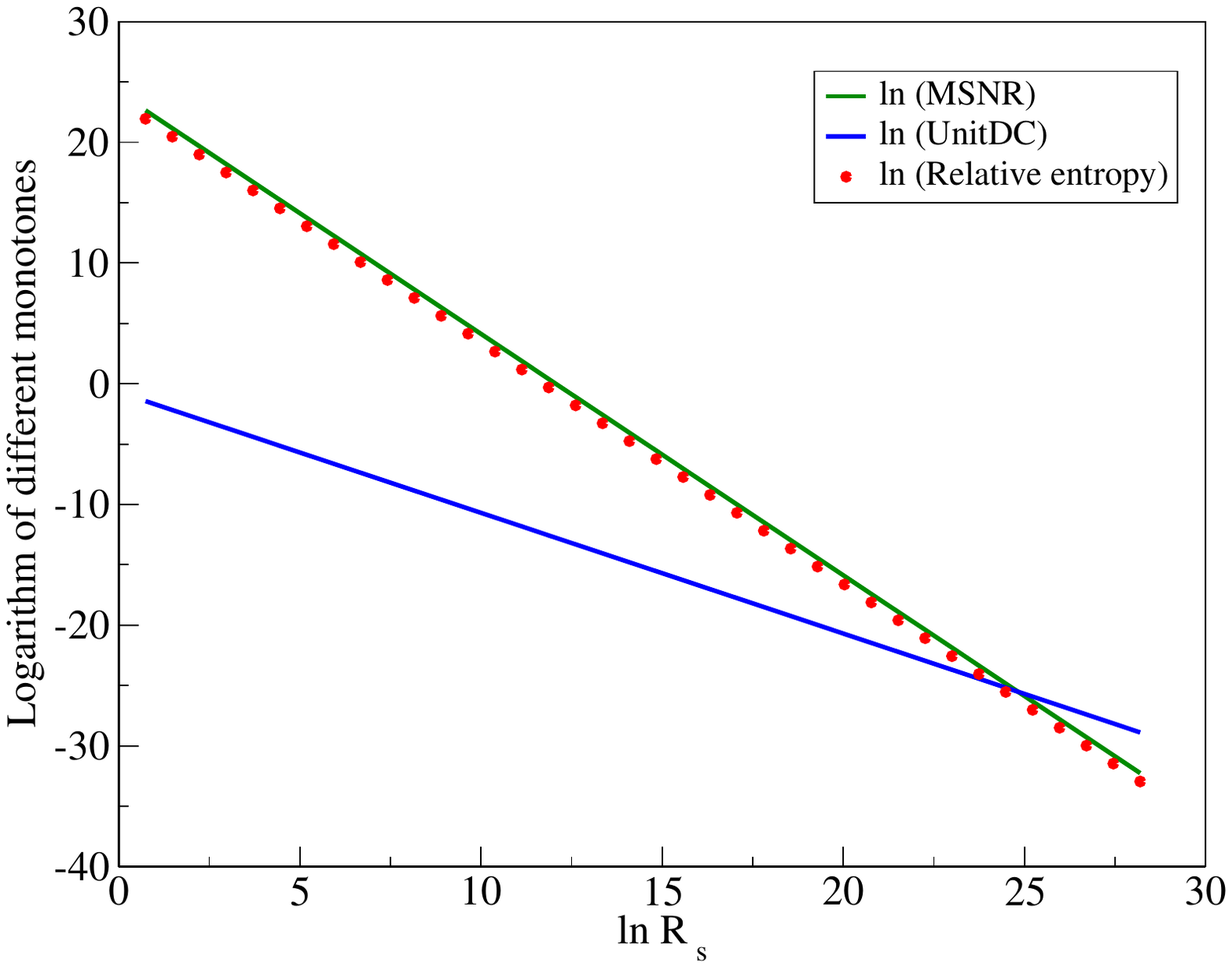}
\caption{Plot of the logarithm (with base $e$) of monotones for a sinusoidal voltage time series, denoted by $\{\mathcal{S}_i\}_{i=0}^{199}$, versus the logarithm of the internal resistance $R_s$, which is monotonically related to the thermal noise at the ambient temperature. We take $\mathcal{S}_i=\sin\left(\frac{\pi}{25}i\right)$ and the noise to be the thermal noise corresponding to its internal resistance $R_s$. The MSNR is given by $\mathrm{MSNR}\left((P(\Svec+\vec{r}_{th}), R_s)\right)=\sum_{i=0}^{199}\sin^2\left(\frac{\pi}{25}i\right)/(800 k_B T \Delta f R_s^2)$ (see Eq. \eqref{form:snr}). The $\unitdc$ is given by $\unitdc\left((P(\Svec+\vec{r}_{th}), R_s)\right)=\frac{1}{200R_s}\sum_{i=0}^{199}\sin^2\left(\frac{\pi}{25}i\right)$ (see Eq. \eqref{eq:val-unit}). The  relative entropy is given by $\mathcal{R}\left((P(\Svec+\vec{r}_{th}), R_s)\right) = \frac{1}{400R_s}\sum_{i=0}^{199} \frac{\sin^2\left(\frac{\pi}{25}i\right) }{4 k_B T \Delta f R_s}$ (see Eq. \eqref{eq:val-rel}). Moreover, we take the pulse duration $\tau=1s$,  $T=300 $ K, $\Delta f=10^3$ MHz, and $R_s$ is taken to vary in the range of $1\Omega$ to $10^{12}\Omega$. Those expressions imply that when $T$ is close to zero (i.e., no noise), the monotones MSNR and relative entropy diverge while $\unitdc$ is finite. This can be seen in the figure by noting that there is a logarithmic scale and the $\unitdc$ values are negative whereas the other two monotones have positive values for small $R_s$ (implying exponential increase as $R_s$ decreases.)}
\label{fig:monotones-comparison}
\end{figure}

We show in Sec.~\ref{sub-distill} below that from many copies of the given source $\vsource$ one can by free operations distill out $\unitdc\vsource$ unit dc sources per original source. $\unitdc$ further has the desirable property of not diverging for finite $R_s$ whereas MSNR does diverge, as exemplified in Fig.~\ref{fig:monotones-comparison}. 

Finally, we note that unitdc can be written in several forms, including 
$\unitdc=\unitdcdef$, where $\mathbb{E}_t$ denotes a time average over the pulse duration $\tau$, $\Delta f$ is the bandwidth under consideration (a setting in an oscilloscope), $k_B$ Boltzmann's constant, $T$ the ambient temperature, and $T_s$ the effective source noise temperature. 

\subsection{Relative entropy between the noisy and the free source}
\label{sub:rel-mon}
Relative entropy-based monotones such as relative entropy of entanglement in the resource theory of entanglement~\cite{Horodecki2009} are widely used in resource theories. Relative entropy measures how different two probability distributions are, therefore, we now introduce and analyze a relative entropy-based measure in our resource theory here.

Let $\mathbb{P}$ and $\mathbb{Q}$ be two probability distributions defined on the same probability space $\xi$ with probability densities $p(x)$ and $q(x)$, respectively, then the relative entropy or Kullback-Leibler divergence between $\mathbb{P}$ and $\mathbb{Q}$ is defined as
\begin{align}
D\left(\mathbb{P}||\mathbb{Q}\right):=\int\mathrm{d}x~ p(x)\ln\frac{p(x)}{q(x)}.
\end{align}
The above expression is finite only when $q(x)=0$ implies $p(x)=0$~\cite{Thomas2006}.
We let 
\begin{align}
\label{eq:rel-ent-new}
\mathcal{R} \vsource := \frac{\tau}{N R_s} D\left(\mathcal{P}|| \mathcal{Q}\right),
\end{align}
where $ \mathcal{P} \equiv p(\vec{\mathcal{V}})$ is the probability distribution for the noisy signal and $\mc{Q} \equiv q(\vec{\mathcal{V}})$ {is the probability distribution for the noise corresponding to the internal resistance $R_s$ at the ambient temperature.}  $\tau$ is again the pulse duration of the source, such that longer pulses can be more valuable than shorter ones. $N$ is the number of time steps in the voltage time series, i.e., $\mathcal{V} = (V_1,\cdots, V_N)^{\dagger}$. Dividing by $N$ counteracts divergence under faster sampling rates (higher $N$). Dividing by $R_s$ is to ensure the monotone decreases under the addition of resistance.
It is immediate from Eq. \eqref{eq:rel-ent-new} that the value of the monotone for a hotter or cooler than the ambient temperature resistor is non-negative.  We now show that $\mathcal{R} \vsource$ is a resource monotone. 

\begin{itemize}
\item {\it Faithfulness:} The relative entropy $D\left(\mathcal{P}|| \mathcal{Q}\right)$ is zero if and only if $\mathcal{P}=\mathcal{Q}$ implying that for thermal noise corresponding to resistance $R_s$ at the ambient temperature, $\mathcal{R} \vsource=0$ and vice versa. Thus, $\mathcal{R} \vsource$ is a faithful resource quantifier. 

\item {\it Monotonicity under free operations:} The orthogonal transformations reversibly map different voltage series with the same 2-norm to each other, thus acting as permutation matrices on the vectors $p(\vec{V})$ whose entries are probabilities of different voltage distributions. Permutation matrices are doubly stochastic (rows and columns sum to 1 and entries are in the range 0 to 1). Moreover, the internal resistance $R_s$, by definition, remains invariant under orthogonal transformations and therefore, $\mc{Q}$ is also invariant.
Now, the data-processing inequality for relative entropy (see Appendix \ref{append:data-process}), which holds for stochastic matrices, applies here. 
\begin{align}
\mathcal{R}\left(\vsource \right) &= \frac{\tau}{NR_s}D\left(\mathcal{P}|| \mathcal{Q}\right)\nonumber\\
&\geq \frac{\tau}{NR_s}D\left(M \mathcal{P}|| M \mathcal{Q}\right)\nonumber\\
&= \frac{\tau}{NR_s}D\left(M \mathcal{P}|| \mathcal{Q}\right)\nonumber\\
&= \mathcal{R}\left((p(\vec{\mathcal{V}'}),R_s) \right),
\end{align}
where $M$ is a column stochastic  matrix and $\vec{\mathcal{V}'}$ is obtained after the orthogonal transformation on $\vec{\mathcal{V}}$. Thus, $\mathcal{R}\vsource$ is nonincreasing under reversible energy-preserving  operations. Further, for any noisy source, adding thermal noise corresponding to some resistance $R$ at ambient temperature does not increase the value of the monotone. The reason is as follows. Note that adding thermal noise will amount to changing $\vec{r}\mapsto \vec{r}+\vec{s}_{th},~\vec{r}_{th}\mapsto \vec{r}_{th}+\vec{s}_{th}~\text{and}~R_s\mapsto R_s+R$, where $\vec{s}_{th}$ is the thermal noise at ambient temperature corresponding to a resistor with resistance $R$. Thus, adding thermal noise is equivalent to a convolution of probability distributions of the signal and noise by a same probability distribution. From~ Appendix \ref{append:convolution} we see that the relative entropy is nonincreasing under convolution. Therefore, based on the above, we conclude that the relative entropy of resource is a faithful resource monotone for our case.
\end{itemize}

\noindent
For the case of an arbitrary noisy signal $\vec{\mathcal{V}}=\Svec + \vec{r}$, where $\Svec$ is a deterministic signal and is defined by
\begin{align}
p(\Svec) &= \prod_i p_i(\Svec_i) = \prod_i \delta(\Svec_i-s_i),
\end{align}
while $\vec{r}$ is satisfying Eq. \eqref{eq:prob-r}, we have
\begin{align}
\label{joint-prob-d1}
p(\vec{\mathcal{V}}) &= \prod_i p_i(V_i)=\prod_i \frac{1}{\sqrt{2\pi\sigma_i^2}} \exp\left[-\frac{(V_i-s_i)^2}{2\sigma_i^2}\right].
\end{align}
Also, let the distribution of the thermal noise corresponding to the internal resistance $R_s$ at the ambient temperature be given by
\begin{align}
\mathcal{Q}(\vec{\mathcal{V}}) &= \prod_i \mathcal{Q}(V_i)= \prod_i  \frac{1}{\sqrt{2 \pi \xi_i^2}} \exp\left[-\frac{V_i^2}{2\xi_i^2}\right].
\end{align}
Then, for such a voltage source, we prove in Appendix \ref{append-proof-rel-ent} that
\begin{align}
\label{eq:val-rel}
\mathcal{R} \vsource= \frac{\tau}{2NR_s}\sum_i \left(\frac{\sigma_i^2+s_i^2}{\xi_i^2}- \ln \left(\frac{\sigma_i^2}{\xi_i^2 }\right)   -1 \right).
\end{align}

\section{Composed sources}
Composed sources are also sources, and they can be given (Thevenin) equivalent circuits with one voltage source and one internal resistance. One reason composed sources are of interest is that we  argue that the $\unitdc\vsource$ corresponds to the fraction of unit dc sources one can gain from many composed copies of the given voltage source. Another is that energy harvesters may have different subcomponents, e.g.\ piezos with different resonance frequencies, as in Ref.~\cite{Li2016}. We first discuss how to compose sources, go on to derive results on how the resource quantifiers behave under the composition of sources and then focus on how many unit dc sources are needed to compose a given source and vice versa how many unit dc sources one can distill from the source.  

\subsection{How to compose sources}
We follow the standard paradigm for composing voltage sources in electronics: composition in series is allowed for any source and composition in parallel is allowed for identical sources. We do not consider what will happen if nonidentical sources are connected in parallel~\footnote{The characters of each source would need to alter in such a scenario in order not to violate Kirchoff's voltage law which implies e.g.\ that two sources connected in parallel should have the same voltage}.

We denote composition in series as $\vsource\circ \vsource$. Under this composition, 
 the voltages and resistances add. It is natural to define the sum of two deterministic signals as the signal of the composite system and similarly the addition of the two noise vectors as the modified noise vector.

\subsection{Resource quantifiers under composition of sources}
We see quickly what the MSNR and unitdc are when composing identical sources.  Note that the sum of two independent centered Gaussian variables each with standard deviation $\sigma$ has standard deviation $\sigma'=\sqrt{2}\sigma$  [since $\sigma'^2=\mathbb{E} (r_1+r_2)^2=\sigma^2+\sigma^2$]. Thus, letting $\circ$ denote composition in series, we have
\begin{align}
\label{def:add-snr}
&\mathrm{MSNR}(\vsource\circ \vsource)\nonumber\\
&:= \frac{1}{2R_s}\frac{\mathbb{E}(|2\Svec|^2)+2\left| \mathbb{E}| \vec{r}|^2 - \mathbb{E}| \vec{r}_{th} |^2\right|}{2\mathbb{E}(|\vec{r}|^2)} \tau\nonumber\\
&=\mathrm{MSNR}\vsource- \frac{1}{2R_s}\frac{\left| \mathbb{E}| \vec{r}|^2 - \mathbb{E}| \vec{r}_{th} |^2\right|}{\mathbb{E}(|\vec{r}|^2)} \tau.
\end{align}
Similarly, for the resource monotone $\unitdc$, we have
\begin{align}
\label{eq:add-unit-dc}
&\unitdc(\vsource\circ \vsource)\nonumber\\
&=\frac{\mathbb{E}|2 \Svec |^2 +2\left| \mathbb{E}| \vec{r}|^2 - \mathbb{E}| \vec{r}_{th}|^2\right|}{2R_s} \tau\nonumber\\
&=2 \unitdc\vsource - \frac{\tau}{R_s}
\left| \mathbb{E}| \vec{r}|^2 - \mathbb{E}| \vec{r}_{th} |^2\right|.
\end{align}
Thus except for the case of $\vec{r}=\vec{r}_{th}$, composing two identical sources lowers the value per source. This is actually to be expected intuitively since the composition in series will lead to some of the random noise cancelling, and is less optimal than firstly rectifying and then composing, as discussed in~Appendix \ref{append:sec-dist}. 

For the case when $\vec{r}=\vec{r}_{th}$, we see that MSNR has the strange feature of not increasing under composition of two identical sources whereas $\unitdc\vsource$ doubles. This again speaks in favor of $\unitdc\vsource$ as a measure of the usefulness of a source.

\subsection{Composing many identical sources}
\label{sub-distill}
When many identical sources are connected in series, the resulting composed source can see the Gaussian noise effectively cancelling due to the law of large numbers. 
Consider $n$ noisy voltage sources, with the same internal resistance $R_s$, given by $\vec{\mathcal{V}} = \vec{\mathcal{S}} +  \vec{r}$ with $p(\vec{\mathcal{V}}) = \prod_i p_i(V_i)$, where $V_i$ are components of $\vec{\mathcal{V}}$ and $p_i(V_i)$ is given by
\begin{align*}
p_i(V_i)= \frac{1}{\sqrt{2\pi \sigma_i^2}}\int \mathrm{Pr}(\mathcal{S}_i=s_i) \exp\left[-\frac{(V_i-s_i)^2}{2\sigma_i^2}\right] \mathrm{d}s_i,
\end{align*}
where $\mathrm{Pr}(\mathcal{S}_i=s_i)$ is the distribution of the signal, which in the deterministic case is just a $\delta$ function. These $n$ noisy sources can be considered as $n$ independent identically distributed vector valued random variables, namely, $\vec{\mathcal{V}}_1, \vec{\mathcal{V}}_2,\cdots,\vec{\mathcal{V}}_n$. Therefore, given any small number $\epsilon>0$ and $\delta>0$, from weak law of large numbers \cite{Thomas2006}, we have for sufficiently large $n$
\begin{align*}
\mathrm{Pr}\left( \left|\frac{1}{n} \sum_{i=1}^n \vec{\mathcal{V}}_i  - \mathbb{E}\left(\vec{\mathcal{V}}\right)\right|\leq \delta \right)\geq 1-\epsilon.
\end{align*}
This means that the source $\frac{1}{n} \sum_{i=1}^n \vec{\mathcal{V}}_i $ is very close to the source $\mathbb{E}\left(\vec{\mathcal{V}}\right)$ in probability. Therefore, we can consider $\frac{1}{n} \sum_{i=1}^n \vec{\mathcal{V}}_i \approx \mathbb{E}\left(\vec{\mathcal{V}}\right)$ in large $n$ limit. Here $\mathbb{E}\left(\vec{\mathcal{V}}\right) = (\mathbb{E}(V_1), \mathbb{E}(V_2),\cdots, \mathbb{E}(V_N))^{\dagger}$ and for all $i=1,\cdots, N$
\begin{align*}
&\mathbb{E}(V_i)\nonumber\\
&=\frac{1}{\sqrt{2\pi \sigma_i^2}}\int V_i \mathrm{d}V_i\int \mathrm{Pr}(\mathcal{S}_i=s_i) \exp\left[-\frac{(V_i-s_i)^2}{2\sigma_i^2}\right] \mathrm{d}s_i \nonumber\\
&=\int \mathrm{d}s_i \mathrm{Pr}(\mathcal{S}_i=s_i) s_i =\mathbb{E}\mathcal{S}_i.
\end{align*}
Thus, $(\overbrace{\vec{\mathcal{V}} + \cdots + \vec{\mathcal{V}}}^n )$ correspond to $n$ $(\mathbb{E}\mathcal{S}_1,\cdots, \mathbb{E}\mathcal{S}_N)^{\dagger}$ voltage vectors, each with internal resistance $R_s$, for large $n$. In particular, when $\Svec$ is a deterministic signal then $n$ copies of $\vsource$ simply corresponds to a source $(n\Svec, nR_s)$, for large $n$.

\subsection{How many unit dc sources in a given source}
We now consider the question of how many unit dc sources are `contained' in a single source. For example, 2s of 1V at 1$\Omega$ internal resistance should intuitively be equivalent to 2 unit dc sources (recall a unit dc source is 1s of 1V with 1$\Omega$ internal resistance at ambient temperature). This process is more generally termed {\em distillation} of standard resources (see, e.g.,~Ref. \cite{Horodecki2009}). It is an appealing property of a resource monotone if it can be interpreted as the number of distillable standard resources. In our case, as discussed above, the natural standard resource is the unit dc source. Thus in our distillation protocol, we are interested in adding $n$ noisy voltage sources and hope to obtain $m$ 1 V dc sources, where $m$ is as large as possible. This may be written as 
\begin{align}
\label{eq:d-transform}
(\overbrace{\vec{\mathcal{V}} + \cdots + \vec{\mathcal{V}}}^n ) ~\xmapsto[\mathrm{Operations}]{\mathrm{Free}}~ (\overbrace{\vec{e} + \cdots + \vec{e} }^m ),
\end{align}
where $\vec{e}$ is 1 V dc voltage source (or voltage time series) with internal resistance $R=1\Omega$ (see Eqs. \eqref{eq:val-one-dc1} and \eqref{eq:val-one-dc}). It is commonly the case that such transformations are identified only in the asymptotic limit of $n\rightarrow \infty$. The law of large number effects can then be employed to cancel out fluctuations away from the desired state in the asymptotic limit as described above. The {\em distillation rate} is thus commonly defined as  $\lim_{n\rightarrow\infty} \frac{m}{n}$. 


We now seek the optimal distillation rate: how many unit dc sources  can we distill from a given source, per copy, via free operations?

Consider firstly, for simplicity, the case of the source having thermal noise at the ambient temperature, plus some signal. Then the $\unitdc$ value of $n$ sources in series has the neat form $nd$ where $d$ is the $\unitdc$ value of a single source. It is possible to convert these $n$ sources in series to $m=nd$ unit dc sources via the free orthogonal operations provided that $m=n \tau\mathbb{E}|\Svec|^2/R_s=nd$ (see Eqs.~\eqref{eq:val-one-dc1} and \eqref{eq:val-one-dc}). This corresponds to the forwards transformation of Eq.~\eqref{eq:d-transform}. Thus one can distill $\unitdc$ units per source via free operations. We give heuristic arguments in~Appendix \ref{append:sec-dist} that, in the reverse  direction, free transformations can convert $m$ unit sources into $nd$ copies of the original source. Thus the unitdc is an achievable and, we believe, optimal rate for sources where the noise is thermal.

We also believe the unitdc is the optimal rate for the case where the sources have Gaussian noise that is not at the ambient temperature. In this case adding the sources in series is not an optimal distillation protocol but rather one should apply a rectifying operation first. This necessitates expanding the set of free operations to include a rectifying operation. We give a possible idealized rectifying operation that can reasonable be called free in~Appendix \ref{append:sec-dist}. It takes free states to free states. Under this operation we show heuristically that $\unitdc (\vsource)$ should indeed be the optimal rate for such sources as well.

\section{Conclusion and future directions}
\label{sec:conc}
We consider transient voltage sources and how to rank them according to their usefulness. We took a so-called resource theory approach to define an interconversion hierarchy. We defined a resource theory for transient voltage sources. We took power-preserving reversible operations, as well as the addition of ambient-temperature Gaussian noise as free interconversion operations. We gave three measures of usefulness that respect the interconversion hierarchy induced by the free operations, and show that some measures currently used do not respect it.  One of our proposed measures, the unitdc, has the interpretation of how many $1$V dc over $1$s, $1\Omega$ internal resistance voltage sources can be distilled per source. The results show that noise that is not from the ambient temperature does carry value and specify how much so. 

We hope this measure may prove useful in comparing different voltage sources fairly and in guiding the optimization of energy harvesting. The results suggest several directions for further development are likely to be fruitful: (i) the quantifiers can be applied to place value on a range of imperfect voltage sources, (ii) the approach can be developed further by extending the set of states (e.g.\ more rich noise models) and free operations (many more converting circuits should be possible), (iii) the choice of free operations to apply for a given conversion is an appealing application of machine-learning techniques, leading to what may be called intelligent energy harvesting, (iv) the efficiency of energy harvesters can be quantified in terms of the amount by which they decrease the resource value of the original, e.g.\ mechanical source, by converting it into electrical output, (v) the law-of-large-numbers approach taken here should more generally be replaced with single-shot approaches similar to what is being done in information theory and statistical mechanics, and the agent's risk-reward profile should be taken into account when assigning value to a resource, and (vi) it will be interesting to apply the same approach to quantum coherent electronics.

\medskip

\begin{acknowledgements}
Parts of this work were carried during the visit of U. S. to SUSTech and U. S. acknowledges the warm hospitality of O. D. at SUSTech. U. S. also acknowledges the support by Polish National Science Center (NCN) (Grant No. 2019/35/B/ST2/01896). O.D. acknowledges support from the National Natural Science Foundation of China (Grants No. 12050410246 and 12005091) and the Shenzhen Science Technology and Innovation Commission (Grant No.20200805101139001). We are grateful for discussions with Masahito Hayashi, Guo Hengyu,  Dario Egloff, and Feiyang Li. We are particularly grateful to Feiyang Liu for all his inputs and discussions. 
\end{acknowledgements}

\bibliography{res-cit}


\appendix

\section{Orthogonal matrices for inter-conversion of certain deterministic voltage sources}
\label{append:ex-orth-matrix}
Given a $1$ V deterministic voltage source for a duration of $\tau=1$s, we can sample it $N$ times in the increments $\delta t = \frac{\tau}{N}=\frac{1}{N}s$. Then this deterministic voltage source can be represented by an $N$ length vector $\vec{V}(N)=(1,\cdots,1)^{\dagger}$. When $N$ is even, we can rewrite $\vec{V}(N)$ as
\begin{align*}
  \vec{V}(N)  =(\overbrace{1,\cdots, 1}^{N/2})^{\dagger}\otimes (1,1)^{\dagger}=\vec{V}(N/2)\otimes \vec{V}(2),
\end{align*}
where $\otimes$ is the Kronecker product. Further, for the even $N$ case, consider another $\sqrt{2}$V deterministic voltage source for a duration of $\frac{\tau}{2}=\frac{1}{2}$s sampled in the intervals of $\delta t = \frac{\tau}{N}=\frac{1}{N}$s. Such a source then can be represented as an $N$ length vector
\begin{align*}
    \vec{U}(N)=\sqrt{2}(\overbrace{1,\cdots,1}^{N/2}, \overbrace{0,\cdots,0}^{N/2})^{\dagger}.
\end{align*}
 It is clear that $\vec{V}(N)^{\dagger} \vec{V}(N) = \vec{U}(N)^{\dagger} \vec{U}(N) = N$. Therefore, there will always exist an orthogonal matrix $O$ such that $\vec{U}(N) = O\vec{V}(N)$. Such an $O$ can be constructed as follows. First, note that using the Hadamard matrix
 \begin{align}
 \label{eq:hada-matrix}
 H=\frac{1}{\sqrt{2}}\begin{pmatrix}1 & 1\\ 1 & -1\end{pmatrix},
 \end{align} 
 which is orthogonal, we get
 \begin{align}
 \label{eq:act-hada-one}
 H\begin{pmatrix}1\\1\end{pmatrix} =\sqrt{2} \begin{pmatrix}1\\0\end{pmatrix}.
 \end{align}
 Now, let us consider a matrix
 \begin{align*}
 O'=\mathbb{I}_{\frac{N}{2}}\otimes H,
 \end{align*} 
 where $\mathbb{I}_{n}$ is the $n\times n$ identity matrix. Using $O'$, we obtain
 \begin{align*}
 \vec{V}'(N)&:=O'\vec{V}(N)\\
 &= \vec{V}(N/2)\otimes \begin{pmatrix}\sqrt{2}\\0\end{pmatrix}\\
 &=\sqrt{2}(\overbrace{1,0}^{},\overbrace{1,0}^{},\cdots, \overbrace{1,0}^{})^{\dagger}.
 \end{align*} 
 It is not difficult to see that after applying a suitable permutation matrix to $\vec{V}'(N)$ we obtain $\vec{U}(N)$. In particular, let the cycle $Q'_{ij}=(ij)$ denote a permutation matrix such that it exchanges indices $i$ and $j$ and leaves all other indices intact. Then the orthogonal matrix
 \begin{align*}
     Q&=  Q'_{N-1,N/2}\cdot Q'_{N-3,N/2-1}\cdot \cdots\cdot Q'_{5,3}\cdot Q'_{3,2}\\
     &=\Pi_{i=1}^{N/2-1}Q'_{2i+1,i+1}
 \end{align*}
gives $\vec{U}(N)=Q\vec{V}'(N) = (Q O')\vec{V}(N)$. Therefore, the desired orthogonal transformation is given by $O=Q O'$.
\\
\\
For $N=2$ case, the desired orthogonal transformation is given by the Hadamard matrix itself (see Eq. \eqref{eq:act-hada-one}). For the $N=4$ case, the matrix $O'$ is given by
\begin{align*}
 O'&=\mathbb{I}_{2}\otimes H=\frac{1}{\sqrt{2}}\begin{pmatrix}
 1 & 1 & 0 & 0 \\
 1 & -1 & 0 & 0\\
 0 & 0 & 1 & 1\\
 0 & 0 & 1 & -1
 \end{pmatrix}.
 \end{align*} 
 The matrix $Q$ is given by
 \begin{align*}
 Q&=Q'_{3,2}=\frac{1}{\sqrt{2}}\begin{pmatrix}
 1 & 0 & 0 & 0 \\
 0 & 0 & 1 & 0\\
 0 & 1 & 0 & 0\\
 0 & 0 & 0 & 1
 \end{pmatrix}.
 \end{align*} 
 The desired orthogonal matrix $O$ is then given by
 \begin{align}
 \label{eq:4d-case-orth}
 Q&=Q O'=\frac{1}{\sqrt{2}}\begin{pmatrix}
 1 & 1 & 0 & 0 \\
 0 & 0 &1 & 1 \\
 1 & -1 & 0 & 0 \\
 0 & 0 & 1 & -1 \\
 \end{pmatrix}.
 \end{align} 
 It is easy to verify that 
 \begin{align*}
 \frac{1}{\sqrt{2}}\begin{pmatrix}
 1 & 1 & 0 & 0 \\
 0 & 0 &1 & 1 \\
 1 & -1 & 0 & 0 \\
 0 & 0 & 1 & -1 \\
 \end{pmatrix}\begin{pmatrix}
 1\\1\\1\\1
 \end{pmatrix}=\sqrt{2}\begin{pmatrix}
 1\\1\\0\\0
 \end{pmatrix}.
 \end{align*} 
 For the case when $N$ is odd, we can obtain the desired orthogonal transformation $O$ satisfying $\vec{v}=O\vec{u}$, where $\vec{u}^{\dagger}\vec{u}=\vec{v}^{\dagger}\vec{v}$, as follows. First note that since both vectors have same length, we can normalize them to make them unit vectors. Then using Gram–Schmidt procedure, we can create two orthonormal basis sets $B_1=\{\vec{u}_i\}_{i=1}^N$ and $B_2=\{\vec{v}_i\}_{i=1}^N$, where $\vec{u}_1=\vec{u}$ and $\vec{v}_1=\vec{v}$. Then the orthogonal matrix $O$ achieving $\vec{v}=O\vec{u}$ is defined by the equation $O\vec{v}_j=\vec{u}_j$ for all $j\in\{1,\cdots,N\}$. Or the matrix elements $O_{ij}$ of $O$ in the basis $B_2$ are given by $\vec{v}_i^{\dagger}O\vec{v}_j=\vec{v}_i^{\dagger}\vec{u}_j$.

\section{Proof of the results in the main text concerning MSNR measure}
\label{append-proofs}
Everywhere, in this Appendix, we assume that the noise in the voltage sources is given by the thermal noise at the ambient temperature, i.e., $\vec{r} =\vec{r}_{th}$ and $\tau=1$.
\subsection{Modified signal-to-noise ratio for noisy sinusoidal signals}
\label{append-snr-thm}
 The MSNR in this case is given by (see Eq. \eqref{MRSNR})
\begin{align}
\label{MRSNR1}
\mathrm{MSNR}\vsource:=\frac{\mathbb{E}|\Svec|^2}{R_s\mathbb{E}|\vec{r}|^2}.
\end{align}
First, let us calculate $\mathbb{E}|\mathcal{S}|^2$.
\begin{align}
\label{eq:av-snr-one}
\mathbb{E}|\Svec|^2&=\frac{1}{N}\int\cdots \int |\Svec|^2 \prod_{k=1}^N  \delta(\mathcal{S}_k - \sin\omega t_k ) \mathrm{d}\mathcal{S}_k\nonumber\\
&=\frac{1}{N}\sum_{k=1}^N\int \mathcal{S}_k^2 \delta(\mathcal{S}_k - \sin\omega t_k ) \mathrm{d}\mathcal{S}_k\nonumber\\
&=\frac{1}{N}\sum_{k=1}^N \sin^2\omega t_k.
\end{align}
Similarly,
\begin{align*}
\mathbb{E}|\vec{r}|^2&=\frac{1}{N}\int\cdots \int |\vec{r}|^2 \prod_{k=1}^N  \frac{1}{\sqrt{2\pi \sigma_k^2}} e^{-\frac{r_k^2}{2\sigma_k^2}}  \mathrm{d}r_k\nonumber\\
&=\frac{1}{N}\sum_{k=1}^N \int r_k^2  \frac{1}{\sqrt{2\pi \sigma_k^2}} e^{-\frac{r_k^2}{2\sigma_k^2}} \mathrm{d}r_k
=\frac{1}{N}\sum_{k=1}^N \sigma_k^2.
\end{align*}
Therefore, from Eq. \eqref{MRSNR1}, we have
\begin{align}
\label{form:snr}
\mathrm{MSNR}\vsource= \frac{1}{R_s}\frac{\sum_{k=1}^N \sin^2\omega t_k}{\sum_{k=1}^N \sigma_k^2}.
\end{align}
Let us further assume the voltage time series contains entries at times $t_k=k \frac{\tau}{N}$. Then, we get
\begin{align*}
 \frac{1}{N}\sum_{k=1}^N \sin^2\omega t_k = \frac{1}{N}\sum_{k=1}^N \sin^2 \left(\frac{\omega k \tau}{N}\right).
\end{align*}
Next, we use the Riemann sum to definite integral formula, which looks like the following:
\begin{align*}
\int_a^b f(x) \mathrm{d}x = \lim_{N\rightarrow \infty} \frac{b-a}{N} \sum_{k=1}^N f\left(a+ \frac{b-a}{N}k\right).
\end{align*}
Now, by choosing $a=0$, $b=\omega \tau$ and $f(x)=\sin^ 2x$ in above formula, we get 
\begin{align*}
\frac{1}{\omega \tau}\int_0^{\omega \tau} \sin^2 x \mathrm{d}x = \lim_{N\rightarrow \infty} \frac{1}{N}\sum_{k=1}^N \sin^2 \left(\frac{\omega k \tau}{N}\right).
\end{align*}
Solving the left-hand side of the above equation, we get 
\begin{align}
\label{eq:int}
\lim_{N\rightarrow \infty} \frac{1}{N}\sum_{k=1}^N \sin^2 \left(\frac{\omega k \tau}{N}\right) &= \frac{1}{\omega \tau}\int_0^{\omega \tau} \sin^2 x \mathrm{d}x\nonumber\\
&=\frac{1}{2}.
\end{align}
Combining the above equation with Eq. \eqref{form:snr}, we have
\begin{align}
\label{eq:simp-form}
\mathrm{MSNR}\vsource= \frac{1}{2R_s}\frac{1}{\lim_{N\rightarrow\infty}\frac{1}{N}\sum_{k=1}^N \sigma_k^2}.
\end{align}
For $\sigma_k=\sigma$ for all $k=1,\cdots, N$, we have 
\begin{align}
\label{eq:simp-form2}
\mathrm{MSNR}\vsource= \frac{1}{2R_s\sigma^2}.
\end{align}

\subsection{Modified signal-to-noise-ratio for voltage sources containing more than one periodic term}
\label{append-snr-thm-broadband}

In the previous subsection, we have consider probability distribution of input voltages with only a single frequency. In the following, we consider the set of all states as before, i.e., $\vec{\mathcal{V}}= \vec{\mathcal{S}} +  \vec{r}$, except the signal $\Svec$  contains multiple frequencies and is distributed according to
\begin{align*}
p(\vec{\mathcal{S}}) =\prod_i  \delta\left(\mathcal{S}_i - \sum_{\omega'} \left(f(\omega')\sin\omega' t_i + g(\omega')\cos\omega' t_i\right)\right),
\end{align*}
for real functions $f(\omega')$ and $g(\omega')$, allowing numerous frequencies. The $\mathrm{MSNR}$ for such a source is again given by
\begin{align}
\label{Eqn:avgbroad}
\mathrm{MSNR_b} &= \frac{1}{R_s}\frac{\mathbb{E}|\Svec|^2}{\mathbb{E}|\vec r|^2}\nonumber\\
&=\frac{\sum_{k=1}^N \left[\sum_{\omega'} \left(f(\omega')\sin\omega' t_k + g(\omega')\cos\omega' t_k\right)\right]^2}{R_s\sum_{k=1}^N \sigma_k^2},
\end{align}
where use of a subscript `$b$' denotes that this $\mathrm{MSNR}$ corresponds to a broadband source.

\bigskip
\noindent 
{\bf Case 1.} When $f(\omega')=\delta(\omega'-\omega)=g(\omega')$. In this case we have following proposition.
\begin{prop}
\label{prop-broad1}
For a voltage time series $\vec{\mathcal{V}}= \vec{\mathcal{S}} +  \vec{r}$  with $p(\vec{\mathcal{S}})=\prod_i  \delta\left(\mathcal{S}_i - \sin\omega t_i -\cos\omega t_i\right)$ and $q(\vec{r})$ satisfying Eq. \eqref{eq:prob-r}, we have 
\begin{align}
\label{eq:sin-cos}
\mathrm{MSNR_b} &= 2~ \mathrm{MSNR} = \frac{1}{R_s} \frac{1}{\lim_{N\rightarrow\infty} \frac{1}{N}\sum_{k=1}^N \sigma_k^2}.
\end{align}
Here $\mathrm{MSNR}$ and $\mathrm{MSNR_b}$ are the signal-to-noise ratios divided by the internal resistance $R_s$ when $p(\vec{\mathcal{S}})=\prod_i  \delta\left(\mathcal{S}_i - \sin\omega t_i \right)$ and $p(\vec{\mathcal{S}})=\prod_i  \delta\left(\mathcal{S}_i - \sin\omega t_i -\cos\omega t_i\right)$, respectively.
\end{prop}

\begin{mproof}
Putting $f(\omega')=\delta(\omega'-\omega)=g(\omega')$ in Eq. \eqref{Eqn:avgbroad}, we have
\begin{align*}
\mathrm{MSNR_b}&= \frac{1}{R_s}\frac{\sum_{k=1}^N \left(\sin\omega t_k + \cos\omega t_k\right)^2}{\sum_{k=1}^N \sigma_k^2}\\
&=\frac{1}{R_s}\frac{N+ \sum_{k=1}^N \sin (2\omega t_k) }{\sum_{k=1}^N \sigma_k^2}.
\end{align*}
Taking $t_k=k \frac{\tau}{N}$, we get
\begin{align*}
 \frac{1}{N}\sum_{k=1}^N \sin(2\omega t_k) = \frac{1}{N}\sum_{k=1}^N \sin \left(\frac{2\omega k \tau}{N}\right).
\end{align*}
Now, using Riemann sum formula, we have
\begin{align}
\frac{1}{\omega \tau}\int_0^{\omega \tau} \sin 2 x ~\mathrm{d}x = \lim_{N\rightarrow \infty} \frac{1}{N}\sum_{k=1}^N \sin \left(\frac{2\omega k \tau}{N}\right).
\end{align}
Solving the left-hand side of the above equation, we get 
\begin{align*}
\lim_{N\rightarrow \infty} \frac{1}{N}\sum_{k=1}^N \sin \left(\frac{2\omega k \tau}{N}\right) = \frac{1}{\omega \tau}\int_0^{\omega \tau} \sin2 x~ \mathrm{d}x=0.
\end{align*}
Thus, we have
\begin{align}
\mathrm{MSNR_b}=\frac{1}{R_s} \frac{1}{\lim_{N\rightarrow\infty} \frac{1}{N}\sum_{k=1}^N \sigma_k^2}.
\end{align}
Using Eq. \eqref{eq:simp-form}, we complete the proof of the proposition.
\end{mproof}

\begin{widetext}

\bigskip
\noindent 
{\bf Case 2.} When $f(\omega')=\delta(\omega'-\omega_1)+ \delta(\omega'-\omega_2)$ and $g(\omega')=0$. In this case we have the  following proposition.
\begin{prop}
\label{prop:broad2}
For a voltage time series $\vec{\mathcal{V}}= (1-g) \vec{\mathcal{S}} + g \vec{r}$  with $p(\vec{\mathcal{S}})=\prod_i  \delta\left(\mathcal{S}_i - \sin\omega_1 t_i -\sin\omega_2 t_i\right)$ and $q(\vec{r})$ satisfying Eq. \eqref{eq:prob-r}, we have 
\begin{align}
\label{snrb3}
\mathrm{MSNR_b}&=\frac{1}{R_s}\frac{1+ \frac{\sin\left(   (\omega_1-\omega_2) \tau \right)}{(\omega_1-\omega_2) \tau} + \frac{\sin\left(   (\omega_1+\omega_2)\tau\right)}{(\omega_1+\omega_2) \tau}}{\lim_{N\rightarrow\infty} \frac{1}{N}\sum_{k=1}^N \sigma_k^2}.
\end{align}
\end{prop}

\begin{mproof}
Putting $f(\omega')=\delta(\omega'-\omega_1)+ \delta(\omega'-\omega_2)$ and $g(\omega')=0$ in Eq. \eqref{Eqn:avgbroad}, we have
\begin{align}
\label{snrb2}
\mathrm{MSNR_b}&= \frac{1}{R_s} \frac{\sum_{k=1}^N \left(\sin\omega_1 t_k + \sin\omega_2 t_k\right)^2}{\sum_{k=1}^N \sigma_k^2}\nonumber\\
&=\frac{1}{R_s}\left[\frac{\sum_{k=1}^N\sin^2 \omega_1 t_k+ \sin^2 \omega_2 t_k  }{\sum_{k=1}^N \sigma_k^2} +\frac{ \sum_{k=1}^N\cos ((\omega_1-\omega_2)t_k) - \cos ((\omega_1+\omega_2)t_k)  }{\sum_{k=1}^N \sigma_k^2}\right].
\end{align}
Now, taking $t_k=k \frac{\tau}{N}$ and using Riemann sum formula, we have
\begin{align*}
&\lim_{N\rightarrow \infty} \frac{1}{N}\sum_{k=1}^N \cos \left(\frac{(\omega_1-\omega_2) k \tau}{N}\right)=\frac{\sin\left(   (\omega_1-\omega_2) \tau\right)}{(\omega_1-\omega_2) \tau}
\end{align*}
and
\begin{align*}
&\lim_{N\rightarrow \infty} \frac{1}{N}\sum_{k=1}^N \cos \left(\frac{(\omega_1+\omega_2) k \tau}{N}\right)=\frac{\sin\left(   (\omega_1+\omega_2) \tau\right)}{(\omega_1+\omega_2) \tau}.
\end{align*}
Using above in Eq. \eqref{snrb2}, we complete the proof of the proposition.

\end{mproof}

\subsection{Modified signal-to-noise ratio for general broadband sources}
\label{append-broad1}
From Eq. \eqref{Eqn:avgbroad}, we have 
\begin{align}
\mathrm{MSNR_b}=\frac{1}{R_s}\frac{\sum_{k=1}^N \left[\sum_{\omega'} \left(f(\omega')\sin\omega' t_k + g(\omega')\cos\omega' t_k\right)\right]^2}{\sum_{k=1}^N \sigma_k^2}.
\end{align}
 Let us define

\begin{align*}
\mathcal{G} &:=  \sum_{k=1}^N \left[\sum_{\omega'} \left(f(\omega')\sin\omega' t_k + g(\omega')\cos\omega' t_k\right)\right]^2\\
&=\sum_{k=1}^N \sum_{\omega,\omega'} \left(f(\omega')\sin\omega' t_k + g(\omega')\cos\omega' t_k\right)\left(f(\omega)\sin\omega t_k + g(\omega)\cos\omega t_k\right)\\
&=\sum_{k=1}^N \sum_{\omega,\omega'} \left(f(\omega') f(\omega) \sin\omega' t_k \sin\omega t_k + g(\omega')f(\omega) \sin\omega t_k \cos\omega' t_k \right) \\
&~~~+\sum_{k=1}^N \sum_{\omega,\omega'} \left(  f(\omega') g(\omega) \sin\omega' t_k  \cos\omega t_k + g(\omega') g(\omega) \cos\omega' t_k \cos\omega t_k  \right)\\
&=\mathcal{G}_1 + \mathcal{G}_2 +\mathcal{G}_3 +\mathcal{G}_4.
\end{align*}
In the following we compute each term divided by $N$ in the limit $N\rightarrow\infty$. 
\begin{align*}
\lim_{N\rightarrow\infty}\frac{1}{N}\mathcal{G}_1&= \lim_{N\rightarrow\infty} \frac{1}{N}\sum_{k=1}^N \sum_{\omega,\omega'} f(\omega') f(\omega) \sin\omega' t_k \sin\omega t_k \\
&=\lim_{N\rightarrow\infty} \frac{1}{N}\frac{1}{2}\sum_{k=1}^N \sum_{\omega,\omega'} f(\omega') f(\omega) \left(\cos\left(\frac{\omega'-\omega) k\tau}{N}\right)- \cos\left(\frac{\omega'+\omega) k\tau}{N}\right) \right) \\
&= \frac{1}{2\tau}\int_0^{\tau} \mathrm{d}x \sum_{\omega,\omega'} f(\omega') f(\omega) \cos(\omega'-\omega) x-\frac{1}{2\tau}\int_0^{\tau} \mathrm{d}x \sum_{\omega,\omega'} f(\omega') f(\omega) \cos(\omega'+\omega) x\\
&= \sum_{\omega,\omega'} f(\omega') f(\omega) \frac{1}{2(\omega'-\omega)\tau} \sin(\omega'-\omega) \tau-\sum_{\omega,\omega'} f(\omega') f(\omega) \frac{1}{2(\omega'+\omega)\tau} \sin(\omega'+\omega) \tau\\
&= \sum_{\omega,\omega'} \frac{f(\omega') f(\omega)}{2} \left(\frac{\sin(\omega'-\omega) \tau}{(\omega'-\omega)\tau}-  \frac{\sin(\omega'+\omega) \tau}{(\omega'+\omega)\tau}\right).
\end{align*}

\begin{align*}
\lim_{N\rightarrow\infty}\frac{1}{N}\mathcal{G}_2&= \lim_{N\rightarrow\infty} \frac{1}{N}\sum_{k=1}^N \sum_{\omega,\omega'} g(\omega') f(\omega) \sin\omega t_k \cos\omega' t_k \\
&=\lim_{N\rightarrow\infty} \frac{1}{N}\frac{1}{2}\sum_{k=1}^N \sum_{\omega,\omega'} g(\omega') f(\omega) \left(\sin\left(\frac{\omega+\omega') k\tau}{N}\right) + \sin\left(\frac{\omega-\omega') k\tau}{N}\right) \right) \\
&= \frac{1}{2\tau}\int_0^{\tau} \mathrm{d}x \sum_{\omega,\omega'} g(\omega') f(\omega) \sin(\omega+\omega') x-\frac{1}{2\tau}\int_0^{\tau} \mathrm{d}x \sum_{\omega,\omega'} g(\omega') f(\omega) \sin(\omega-\omega') x\\
&= \sum_{\omega,\omega'} \frac{g(\omega') f(\omega)}{2} \left(\frac{1-\cos(\omega+\omega') \tau}{(\omega+\omega')\tau} + \frac{1-\cos(\omega-\omega') \tau}{(\omega-\omega')\tau} \right)\\
&= \sum_{\omega,\omega'} \frac{g(\omega') f(\omega)}{2} \left(\frac{1-\cos(\omega'+\omega) \tau}{(\omega'+\omega)\tau} - \frac{1-\cos(\omega'-\omega) \tau}{(\omega'-\omega)\tau} \right).
\end{align*}

\begin{align*}
\lim_{N\rightarrow\infty}\frac{1}{N}\mathcal{G}_3&= \lim_{N\rightarrow\infty} \frac{1}{N}\sum_{k=1}^N \sum_{\omega,\omega'} f(\omega') g(\omega) \sin\omega' t_k \cos\omega t_k \\
&=\lim_{N\rightarrow\infty} \frac{1}{N}\frac{1}{2}\sum_{k=1}^N \sum_{\omega,\omega'} f(\omega') g(\omega) \left(\sin\left(\frac{\omega'+\omega) k\tau}{N}\right) + \sin\left(\frac{\omega'-\omega) k\tau}{N}\right) \right) \\
&= \frac{1}{2\tau}\int_0^{\tau} \mathrm{d}x \sum_{\omega,\omega'} f(\omega') g(\omega) \sin(\omega'+\omega) x-\frac{1}{2\tau}\int_0^{\tau} \mathrm{d}x \sum_{\omega,\omega'} g(\omega') f(\omega) \sin(\omega'-\omega) x\\
&= \sum_{\omega,\omega'} \frac{f(\omega') g(\omega)}{2} \left(\frac{1-\cos(\omega'+\omega) \tau}{(\omega'+\omega)\tau} + \frac{1-\cos(\omega'-\omega) \tau}{(\omega'-\omega)\tau} \right).
\end{align*}

\begin{align*}
\lim_{N\rightarrow\infty}\frac{1}{N}\mathcal{G}_4&= \lim_{N\rightarrow\infty} \frac{1}{N}\sum_{k=1}^N \sum_{\omega,\omega'} g(\omega') g(\omega) \cos\omega' t_k \cos\omega t_k \\
&=\lim_{N\rightarrow\infty} \frac{1}{N}\frac{1}{2}\sum_{k=1}^N \sum_{\omega,\omega'} g(\omega') g(\omega) \left(\cos\left(\frac{\omega'+\omega) k\tau}{N}\right)+ \cos\left(\frac{\omega'-\omega) k\tau}{N}\right) \right) \\
&= \frac{1}{2\tau}\int_0^{\tau} \mathrm{d}x \sum_{\omega,\omega'} g(\omega') g(\omega) \cos(\omega'+\omega) x+\frac{1}{2\tau}\int_0^{\tau} \mathrm{d}x \sum_{\omega,\omega'} g(\omega') g(\omega) \cos(\omega'-\omega) x\\
&= \sum_{\omega,\omega'} \frac{g(\omega') g(\omega)}{2} \left(\frac{\sin(\omega'+\omega) \tau}{(\omega'+\omega)\tau} +  \frac{\sin(\omega'-\omega) \tau}{(\omega'-\omega)\tau}\right).
\end{align*}

Combining the above equations, we get

\begin{align*}
\lim_{N\rightarrow\infty}\frac{1}{N} \mathcal{G} &= \sum_{\omega,\omega'} \left(\frac{g(\omega') g(\omega)- f(\omega') f(\omega)}{2}\right)\frac{\sin(\omega'+\omega) \tau}{(\omega'+\omega)\tau} +\left(\frac{g(\omega') g(\omega)+ f(\omega') f(\omega)}{2}\right)\frac{\sin(\omega'-\omega) \tau}{(\omega'-\omega)\tau}\\
&+ \sum_{\omega,\omega'} \left(\frac{g(\omega') f(\omega)+ f(\omega') g(\omega)}{2}\right)\frac{1-\cos(\omega'+\omega) \tau}{(\omega'+\omega)\tau}  +\left( \frac{f(\omega') g(\omega)- g(\omega') f(\omega)}{2}\right)\frac{1-\cos(\omega'-\omega) \tau}{(\omega'-\omega)\tau}.
\end{align*}
Let us consider a special case when $\omega$ takes values from the set $\{\Omega, 2\Omega,\cdots\}$ and $\sum_{\omega}\rightarrow \sum_{m=1}^{\infty}$. Further, choose $f(\omega)\rightarrow \tilde{f}(m)= 0$ and $g(\omega)\rightarrow \tilde{g}(m)= 1/m$. This arises in the following Fourier series:
\begin{align}
-\ln\left(2\sin \frac{|\Omega \tau|}{2} \right) = \sum_{m=1}^\infty \frac{\cos m \Omega \tau}{m}.
\end{align}
Then,
\begin{align*}
\lim_{N\rightarrow\infty}\frac{1}{N} \mathcal{G} &= \sum_{\omega,\omega'} \frac{g(\omega') g(\omega)}{2}\left(\frac{\sin(\omega'-\omega) \tau}{(\omega'-\omega)\tau}+\frac{\sin(\omega'+\omega) \tau}{(\omega'+\omega)\tau}\right) \\
&= \sum_{n,m=1}^\infty \frac{1}{2 m n}\left(\frac{\sin(m-n)\Omega \tau}{(m-n)\Omega \tau}+\frac{\sin(m+n)\Omega \tau}{(m+n)\Omega \tau}\right).
\end{align*}

Thus, we have
\begin{align*}
&\mathrm{MSNR_b}=\frac{1}{R_s}\frac{\sum_{n,m=1}^\infty \frac{1}{2 m n}\left(\frac{\sin(m-n)\Omega \tau}{(m-n)\Omega \tau}+\frac{\sin(m+n)\Omega \tau}{(m+n)\Omega \tau}\right)}{\lim_{N\rightarrow \infty}\frac{1}{N}\sum_{k=1}^N \sigma_k^2}.
\end{align*}

\section{Computation of \texorpdfstring{$\unitdc$}{} for various voltage sources}
\label{append-unitdc-proofs}
In this Appendix again, we assume that the noise in the voltage sources is given by the thermal noise at the ambient temperature, i.e., $\vec{r} =\vec{r}_{th}$ and $\tau=1$. For a voltage time series $\vec{\mathcal{V}}= \vec{\mathcal{S}} +  \vec{r}$  with arbitrary distribution of the signal part and noise distribution $q(\vec{r})$ satisfying Eq. \eqref{eq:prob-r}, the $\unitdc$ and MSNR are related simply as follows
\begin{align}
    \unitdc\vsource = \mathrm{MSNR}\vsource \left(\frac{1}{N}\sum_{k=1}^N\sigma_k^2\right).
\end{align}
Let us consider the broadband case, where signal contains multiple frequencies and is distributed according to
\begin{align*}
p(\vec{\mathcal{S}}) =\prod_i  \delta\left(\mathcal{S}_i - \sum_{\omega'} \left(f(\omega')\sin\omega' t_i + g(\omega')\cos\omega' t_i\right)\right).
\end{align*}
In this case, we have
\begin{align}
    \unitdc_b =\frac{\sum_{k=1}^N \left[\sum_{\omega'} \left(f(\omega')\sin\omega' t_k + g(\omega')\cos\omega' t_k\right)\right]^2}{N R_s},
\end{align}
where the subscript `$b$' again denotes the broadband case. In the following, we present $\unitdc$ for various choices of the frequencies in the signal part.
\\
{\bf Case 1.} When $f(\omega')=\delta(\omega'-\omega)$ and $g(\omega')=0$, i.e., when there is only single sinusoidal term in the signal. In this case, using Eq. \eqref{eq:av-snr-one}, the $\unitdc$ is given by
\begin{align}
    \unitdc =\frac{1}{N R_s}\sum_{k=1}^N \sin^2\omega t_k.
\end{align}
When taking $N\rightarrow\infty$ limit, we get
\begin{align}
    \unitdc =\frac{1}{2 R_s}.
\end{align}
\\
{\bf Case 2.} When $f(\omega')=\delta(\omega'-\omega)=g(\omega')$. In this case, using Eq. \eqref{eq:sin-cos}, we have (in $N\rightarrow\infty$ limit)
\begin{align}
\unitdc_b &= 2~ \unitdc= \frac{1}{R_s}.
\end{align}
\\
{\bf Case 3.} When $f(\omega')=\delta(\omega'-\omega_1)+\delta(\omega'-\omega_2)$ and $g(\omega')=0$. In this case, using Eq. \eqref{snrb3}, we have (in $N\rightarrow\infty$ limit)
\begin{align}
\unitdc_b =\frac{1}{R_s}\left( 1+ \frac{\sin\left(   (\omega_1-\omega_2) \mc{T}\right)}{(\omega_1-\omega_2) \mc{T}} + \frac{\sin\left(   (\omega_1+\omega_2) \mc{T}\right)}{(\omega_1+\omega_2) \mc{T}}\right).
\end{align}
Here $\mc{T}$ can be time period of any of the signal component or any other constant. Similarly, one can write $\unitdc$ for the most general case following Sec. \ref{append-broad1}.

\section{\texorpdfstring{Data processing inequality for relative entropy and proof of Eq. \eqref{eq:val-rel}}{}}
\label{append:data-process}

\subsection{Data processing inequality for relative entropy}
Let us consider two probability distributions $\mathcal{P}=\{\mathcal{P}_i\}$ and $\mathcal{Q}=\{\mathcal{Q}_i\}$ and $M$ a column stochastic matrix. Let $\mathcal{P}'=M\mathcal{P}$ and $\mathcal{Q}'=M\mathcal{Q}$. Now,
\begin{align}
D(\mathcal{P}'||\mathcal{Q}') &= \sum_i \mathcal{P}'_i \ln \frac{\mathcal{P}'}{\mathcal{Q}'}\nonumber\\
&=\sum_i \sum_j M_{ij} \mathcal{P}_j \ln \frac{\sum_k M_{ik} \mathcal{P}_k}{\sum_k M_{ik} \mathcal{Q}_k}\nonumber\\
&\leq \sum_i \sum_j M_{ij} \mathcal{P}_j \ln \frac{\ M_{ij} \mathcal{P}_j}{ M_{ij} \mathcal{Q}_j}\nonumber\\
&=  \sum_j \left(\sum_i M_{ij}\right) \mathcal{P}_j \ln \frac{\ \mathcal{P}_j}{  \mathcal{Q}_j}\nonumber\\
&=D(\mathcal{P}||\mathcal{Q}).
\end{align}
In the above we  use Log-sum inequality \cite{Thomas2006}, which states that
\begin{align}
\sum_i a_i \ln \frac{a_i}{b_i} \geq \left(\sum_i a_i\right) \ln \frac{\sum_j a_j}{\sum_j  b_j}.
\end{align}
Therefore, if $M$ is a column stochastic matrix, then
\begin{align}
D(M\mathcal{P}||M\mathcal{Q}) \leq D(\mathcal{P}||\mathcal{Q}).
\end{align}

\subsection{Relative entropy under convolution of probability distributions}
\label{append:convolution}
Let us consider two probability distributions $p$ and $q$ corresponding to some random variable $X$. Let $Y$ be another random variable with probability density $r$. Then the probability densities corresponding to $X+Y:=Z$ will be given by $P:=p*r$ and $Q:=q*r$ for the cases when $X$ is distributed by $p$ and $q$, respectively. In particular,
\begin{align*}
    P(z)=\int p(x) r(z-x)\mathrm{d}x,~~~~ Q(z)=\int q(x) r(z-x)\mathrm{d}x.
\end{align*}
Now,
\begin{align*}
 D(P||Q)&=\int \mathrm{d}z P(z)\ln\frac{P(z)}{Q(z)}\\ 
 &=\int \mathrm{d}z \left(\int  p(x) r(z-x)\mathrm{d}x\right)\ln\frac{\left(\int  p(x) r(z-x)\mathrm{d}x\right)}{\left(\int  q(x) r(z-x)\mathrm{d}x\right)}\\ 
 &\leq \int \mathrm{d}z \int\mathrm{d}x~ p(x) r(z-x) \ln\frac{ p(x) }{q(x)}\\ 
 &= \int\mathrm{d}x~ p(x) \ln\frac{ p(x) }{q(x)}\\ 
 &=D(p||q),
\end{align*}
where the inequality in line $3$ comes from the continuous analog of the Log-sum inequality~\cite{Thomas2006}. Thus, the relative entropy is nonincreasing under the convolution of probability distributions.

\subsection{\texorpdfstring{Proof of Eq. \eqref{eq:val-rel}}{}}
\label{append-proof-rel-ent}
The relative entropy between the probability distributions of our interest is given by
\begin{align}
\label{eq:first}
D\left(\mathcal{P}|| \mathcal{Q}\right)
 &=\int\cdots\int \mathrm{d}V_1\cdots \mathrm{d}V_N \prod_i p_i(V_i) \ln\left(\frac{\prod_j p_j(V_j)}{\prod_j \mathcal{Q}(V_j)}\right)\nonumber\\
&=\int\cdots\int \mathrm{d}V_1\cdots \mathrm{d}V_N \prod_i p_i(V_i) \sum_j \ln\left(\frac{p_j(V_j)}{ \mathcal{Q}(V_j)}\right)\nonumber\\
&=\sum_i \int \mathrm{d}V_i p_i(V_i)  \ln\left(\frac{p_i(V_i)}{ \mathcal{Q}(V_i)}\right).
\end{align}
Note that, we have
\begin{align}
p_i(V_i)=\frac{1}{\sqrt{2\pi\sigma_i^2}}\exp\left[-\frac{(V_i-s_i)^2}{2\sigma_i^2}\right],
\end{align}
and
\begin{align}
\mathcal{Q}(V_i) =  \frac{1}{\sqrt{2 \pi \xi_i^2}} \exp\left[-\frac{V_i^2}{2\xi_i^2}\right].
\end{align}
Then
\begin{align}
\label{eq:re1}
 \int \mathrm{d}V_i p_i(V_i)  \ln\left(\frac{p_i(V_i)}{ \mathcal{Q}(V_i)}\right) &=\ln \left(\frac{\xi_i}{\sigma_i}\right)  \int \mathrm{d}V_i    p_i(V_i)   + \int \mathrm{d}V_i p_i(V_i) \left[ \frac{V_i^2}{2\xi_i^2} -  \frac{(V_i-s_i)^2}{2\sigma_i^2}\right]   \nonumber\\
&=\ln \left(\frac{\xi_i}{\sigma_i}\right) +I_1-I_2,
\end{align}
where
\begin{align}
\label{eq:re2}
 I_2&=\frac{1}{\sqrt{2\pi \sigma_i^2}}\int \mathrm{d}V_i    \frac{(V_i-s_i)^2}{2\sigma_i^2} \exp\left[-\frac{(V_i-s_i)^2}{2\sigma_i^2}\right]=\frac{1}{2},
\end{align}
and
\begin{align}
\label{eq:third}
 I_1 &=  \frac{1}{\sqrt{2\pi\sigma_i^2}}\int \mathrm{d}V_i   \frac{V_i^2}{2\xi_i^2} \exp\left[-\frac{(V_i-s_i)^2}{2\sigma_i^2}\right]\nonumber\\
 &=   \frac{1}{\sqrt{\pi}}\int \mathrm{d}y   \frac{\left(\sqrt{2\sigma_i^2} y +  s_i\right)^2}{2\xi_i^2}e^{-y^2} \nonumber\\
 &= \frac{\sigma_i^2}{\sqrt{\pi} \xi_i^2} \int \mathrm{d}y ~ y^2 e^{-y^2}    + \frac{1}{\sqrt{\pi}} \frac{s_i^2}{2\xi_i^2} \int \mathrm{d}y   e^{-y^2}  \nonumber\\
 &=  \frac{\sigma_i^2}{2\xi_i^2}   + \frac{s_i^2}{2\xi_i^2}.
\end{align}
Combining Eqs. \eqref{eq:first}, \eqref{eq:re1}, \eqref{eq:re2}, and \eqref{eq:third}, we get
\begin{align}
\label{eq:minim}
D\left(\mathcal{P}|| \mathcal{Q}\right)
 &=\frac{1}{2}\sum_{i=1}^{N} \left(\frac{\sigma_i^2+s_i^2}{\xi_i^2}- \ln \left(\frac{\sigma_i^2}{\xi_i^2 }\right)   -1 \right).
\end{align}
Therefore, the relative entropy of the resource is given by
\begin{align}
\label{eq:av-rel-ent-arb}
\mathcal{R} \vsource =\frac{\tau}{2NR_s}\sum_{i=1}^{N} \left(\frac{\sigma_i^2+s_i^2}{\xi_i^2}- \ln \left(\frac{\sigma_i^2}{\xi_i^2 }\right)   -1 \right).
\end{align}

In the previous paragraph, we considered signals $\Svec$ with arbitrary distributions, which are given by $p(\mc{S}_i) = \delta(\mc{S}_i-s_i)$. In the general broadband case, the signal $\Svec$  contains multiple frequencies and is distributed according to
\begin{align*}
p(\vec{\mathcal{S}}) =\prod_{i=1}^{N}  \delta\left(\mathcal{S}_i - \sum_{\omega'} \left(f(\omega')\sin\omega' t_i + g(\omega')\cos\omega' t_i\right)\right),
\end{align*}
allowing numerous frequencies. In this case, using Eq. \eqref{eq:av-rel-ent-arb}, we have 
\begin{align}
\mathcal{R}_b&= \frac{\tau}{2NR_s}\sum_{i=1}^{N} \left(\frac{\sigma_i^2+\left[\sum_{\omega'} \left(f(\omega')\sin\omega' t_i + g(\omega')\cos\omega' t_i\right)\right]^2}{\xi_i^2}- \ln \left(\frac{\sigma_i^2}{\xi_i^2 }\right)   -1 \right),
\end{align}
where use of a subscript `$b$' denotes the broadband source.

\section{Addition of two non-identical noisy sources}

We now give an example of adding to two {\em nonidentical} noisy sources. Let us consider two voltage sources $(P(\vec{\mathcal{V}})^{(1)}, R_1)$ and $(P(\vec{\mathcal{V}})^{(2)}, R_2)$ as given by $
\vec{\mathcal{V}}^{(\mu)}=\vec{\mathcal{S}}^{(\mu)} + \vec{r}^{(\mu)}
$, where $\mu=1,2$ and $\vec{\mathcal{S}}^{(\mu)}$ is distributed according to
\begin{align}
p\left(\vec{\mathcal{S}}^{(\mu)}\right) =\prod_i  \delta\left(\mathcal{S}^{(\mu)}_i - \sin\omega^{(\mu)} t_i \right),
\end{align}
while $\vec{r}^{(\mu)}$ is distributed according to
\begin{align}
q\left(\vec{r}^{(\mu)}\right) =\prod_i  \frac{1}{\sqrt{2\pi \sigma^{(\mu)2}_i}} \exp\left[-r^{(\mu)2}_i/2\sigma^{(\mu)2}_i\right].
\end{align}

Here $\vec{r}^{(\mu)}$ is considered as thermal noise at ambient temperature. If we add these sources in series then the combined source will have voltage given by $\vec{\mathcal{V}}=\vec{\mathcal{S}} +  \vec{r}$, where $\vec{\mathcal{S}}=\vec{\mathcal{S}}^{(1)} + \vec{\mathcal{S}}^{(2)}$ and $\vec{r}=\vec{r}^{(1)} + \vec{r}^{(2)}$. Moreover the internal resistance will become $R_1+R_2$.  From the convolution theorem, the probability distribution of $ \vec{\mathcal{S}}$ is given by $p( \vec{\mathcal{S}}) = \prod_i p(\mathcal{S}_i)$, where
\begin{align*}
p(\mathcal{S}_i)
&= \int^{\infty}_{-\infty} \mathrm{d}x_i~ \delta\left( \mathcal{S}_i -x_i- \sin\omega^{(2)} t_i \right)   \delta\left(x_i - \sin\omega^{(1)} t_i \right) \nonumber\\
&= \delta\left( \mathcal{S}_i -\sin\omega^{(1)} t_i- \sin\omega^{(2)} t_i \right).
\end{align*}
Similarly, the probability distribution of $ \vec{r}$ is given by $q( \vec{r}) = \prod_i q(r_i)$, where
\begin{align*}
q(r_i) = \frac{1}{\sqrt{2\pi \sigma_i^2}}  \exp\left[-\frac{r_i^2}{2\sigma_i^2}  \right].
\end{align*}
and $\sigma_i^2=\sigma^{(1)2}_i +\sigma^{(2)2}_i$. From Eq. \eqref{snrb3} of previous section, we get
\begin{align}
\label{eq:add-two-source}
&\mathrm{MSNR}\left( (p(\vec{\mathcal{V}})^{(1)}, R_1)\circ(p(\vec{\mathcal{V}})^{(2)}, R_2) \right)=\frac{1}{(R_1+R_2)}\frac{1+ \frac{\sin\left(   (\omega^{(1)}-\omega^{(2)}) T\right)}{(\omega^{(1)}-\omega^{(2)}) T} + \frac{\sin\left(   (\omega^{(1)}+\omega^{(2)}) T\right)}{(\omega^{(1)}+\omega^{(2)}) T}}{\lim_{N\rightarrow\infty} \frac{1}{N}\sum_{k=1}^N \left(\sigma^{(1)2}_k +\sigma^{(2)2}_k\right)}.
\end{align}

Similarly, one can compute $\unitdc$ for such a combined source and we have
\begin{align}
\unitdc \left( (p(\vec{\mathcal{V}})^{(1)}, R_1)\circ(p(\vec{\mathcal{V}})^{(2)}, R_2) \right)&=\frac{\mathbb{E}| \Svec |^2}{(R_1+R_2)}\nonumber\\
&=\frac{\sum_{k=1}^N \left(\sin\omega^{(1)} t_k + \sin\omega^{(2)} t_k\right)^2}{N(R_1+R_2)}.
\end{align}

For the case of relative entropy, note that $\vec{\mathcal{V}}=\vec{\mathcal{S}}^{(1)} + \vec{r}^{(1)}+ \vec{\mathcal{S}}^{(2)}+\vec{r}^{(2)}$ is distributed as 
\begin{align*}
p(V_i) = \prod_i \frac{1}{\sqrt{2\pi \sigma_i^2}}  \exp\left[\frac{\left(V_i-\sin\omega^{(1)} t_i- \sin\omega^{(2)} t_i\right)^2}{2\sigma_i^2}  \right],
\end{align*}
where $\sigma_i^2=\sigma^{(1)2}_i +\sigma^{(2)2}_i$. Let the distribution of the thermal noise corresponding to the internal resistance $R_a$ ($a=1,2$) at the ambient temperature be given by
\begin{align*}
\mathcal{Q}_a(\vec{\mathcal{V}})= \prod_i  \frac{1}{\sqrt{2 \pi \xi_i^{(a)2}}} \exp\left[-\frac{V_i^2}{2\xi_i^{(a)2}}\right].
\end{align*}
Then, the distribution of the thermal noise corresponding to the internal resistance $R_1+R_2$ at the ambient temperature be given by
\begin{align*}
\mathcal{Q}(\vec{\mathcal{V}})= \prod_i  \frac{1}{\sqrt{2 \pi \xi_i^{2}}} \exp\left[-\frac{V_i^2}{2\xi_i^{2}}\right],
\end{align*}
where $\xi_i^{2}=\xi_i^{(1)2}+\xi_i^{(2)2}$. Now, using Eq. \eqref{eq:av-rel-ent-arb}, and $\tau=1s$, we have
\begin{align}
\mathcal{R} \left( (p(\vec{\mathcal{V}})^{(1)}, R_1)\circ(p(\vec{\mathcal{V}})^{(2)}, R_2) \right) =\frac{1}{2N(R_1+R_2)}\sum_{i=1}^{N} \left(\frac{\sigma_i^2+(\sin\omega^{(1)} t_i+ \sin\omega^{(2)}t_i)^2}{\xi_i^2}- \ln \left(\frac{\sigma_i^2}{\xi_i^2 }\right)   -1 \right),
\end{align}
where $\sigma_i^2=\sigma^{(1)2}_i +\sigma^{(2)2}_i$ and $\xi_i^{2}=\xi_i^{(1)2}+\xi_i^{(2)2}$.

\end{widetext}

\section{Distillation and formation of \texorpdfstring{$\unitdc$}{} from noisy voltage sources}
\label{append:sec-dist}
In this section we discuss, in a mixture of rigorous and heuristical arguments, how much $\unitdc$ one can distill from noisy sources. 

\subsection{Distillation of \texorpdfstring{$\unitdc$}{} via rectification of many copies}
How can one distill $\unitdc$ from, e.g., a hot resistor, and how much can be distilled? A hot resistor is a case of the Gaussian noise having  higher variance than that corresponding to the ambient temperature:
\begin{equation}
\label{eq:hotnoise}
\mathbb{E}|\vec{r}|^2>\mathbb{E} |\vec{r}_{T_{amb}}|^2=\mathbb{E} |\vec{r}_{th}|^2.
\end{equation} 
We can say sources respecting Eq. \eqref{eq:hotnoise} have `hot' noise. We now initially assume we have a hot noise source rather than a cold noise source. We let $\vec{r}$ denote Gaussian random variables and $\Svec$ deterministic ones so that a source is written as $(\Svec+\vec{r},R_s)$. For example, a resistor with resistance $R_s$ and temperature $T_H$ is then written as 
$(0+\vec{r}_{T_H},R_s)$. 

How can this distillation of unit dc sources be achieved? Orthogonal transformations cannot do the job, since the source voltage is randomly positive and negative and orthogonal transformations by inspection cannot turn that into a deterministically positive number.  For example in three dimensions one can visualize that if a vector might be pointing in any quadrant--the agent does not know which one--no orthogonal transformation chosen by the agent is guaranteed to make it point into the positive quadrant.

In practice people use {\em rectifiers} (e.g.\ diode bridges) to achieve guaranteed positive voltage output. Let us therefore try to define an idealized rectifier that could in principle be allowed. For a voltage source $V$, it should satisfy two conditions,  (i) rectification, i.e., $V\mapsto |V|-V_b$ where $V_b$ is the penalty associated with potential barriers in the diodes, (ii) take an ambient thermal source to another ambient thermal source (the second law of thermodynamics demands this, assuming the diode is not changed by the operation~\cite{Liu2018}). Real diodes do have internal resistance but for simplicity let us firstly see if we can satisfy the two conditions without such resistance. We thus define the source resistance to be invariant, $R_s\mapsto R_s$, where the arrow again indicates the modification of the source by the addition of the diode.

We can consider defining an idealized rectifier operation acting on hot-noise sources (see Eq. \eqref{eq:hotnoise}) as 
\begin{equation}
\label{eq:rectifier}
p(V)~ \mapsto~ \mathcal{N}( |\mathbb{E} |V|^2-\mathbb{E} |\vec{r}_{th}|^2|^{\frac{1}{2}}, \left(\mathbb{E}| \vec{r}_{th}|^2\right)^{\frac{1}{2}}),
\end{equation}
and $R_s$ being unmodified. $\mathcal{N}(a,b)$ here means normal distribution with mean $a$ and standard deviation $b$. $\mathbb{E} |\vec{r}_{th}|^2|^{\frac{1}{2}}$ can be interpreted as the penalty term, the lowering of the voltage due to the rectification. Condition (i) of rectification is then satisfied in the regime where thermal noise is insignificant  ($V^2\gg \mathbb{E}| \vec{r}_{th}|^2$), since in that regime the idealized rectification would take  $V\mapsto |V|$. Moreover the output voltage is generally lower than $|V|$, even though it does not have the exact form $|V|-V_b$ (an expression which anyway should not be taken too seriously as it could be negative). Condition (ii) is moreover satisfied since 
$$\mathcal{N}(0, \! |\mathbb{E}| \vec{r}_{th}|^2|^{\frac{1}{2}})~\mapsto~ \mathcal{N}( | \mathbb{E} |\vec{r}_{th}|^2- \mathbb{E} |\vec{r}_{th}|^2|^{\frac{1}{2}}, \! \left(\mathbb{E}| \vec{r}_{th}|^2\right)^{\frac{1}{2}}).$$

Let us then accept Eq. \eqref{eq:rectifier} as a possibly, in principle, realizable rectifier and go on to use this idealized rectifier operation and see how many unit dc sources we can distill out of a transient source. 

Given a source $(\Svec+\vec{r},R_s)$ we then see that the rectifier turns the voltage distribution into  
\begin{eqnarray*}
\mathcal{N}(|\mathbb{E} |\Svec|^2 +\mathbb{E} |\vec{r}|^2 - \mathbb{E}
|\vec{r}_{th}|^2|^{\frac{1}{2}}),\mathbb{E} |\vec{r}_{th}|^2).
\end{eqnarray*}

Now we allow for one more tool to achieve the distillation: composition of many independent identical copies of the source. Consider many, $n$,  such sources in series. By the law of large numbers, the total voltage is then approximately deterministic with value $$V_{out}=n|\mathbb{E }|\Svec|^2 + \mathbb{E}|\vec{r}|^2-\mathbb{E}|\vec{ r}_{th}|^2|^{\frac{1}{2}}.$$

How many unit dc sources is this equivalent to? A unit dc source can we written as $(\vec{S}+\vec{r}, R_s) =(1+r_{th}, 1)$. Here we have instead $(V_{out}+r_{th}, R_s)$. More generally, suppose we want to create $(c,R)$, a constant voltage $c$ and internal resistance $R$ from unit dc sources meaning $(1,1)$. Recall that two voltage sources with identical emfs connected in parallel have a net emf equivalent to one emf source, however, the net internal resistance is less ($\frac{1}{R_{net}}=\sum_i\frac{1}{R_i}$).  Consider the following protocol: connect in series $c$ sources each with resistance $R/c$ and voltage $1$. We can create each such elementary source by putting $c/R$ unit dc sources in parallel. Thus we would end up using $\frac{c^2}{R}$ unit dc sources to mimic the target source. One reason to believe this is optimal is that it is the same number of unit dcs that would give the maximum power, with the maximization over resistive loads for the target source ($\frac{c^2}{R}$). We therefore conjecture and henceforth assume (we call it a conjecture because we did not describe a protocol for breaking an arbitrary source into unit dc sources) that $(c,R)$ can in principle be converted to and from $\frac{c^2}{R}$ unit dc sources. 

Accordingly, the source $(V_{out}, R_s)$ composed out of hot-noise sources is equivalent to $V_{out}^2/(nR_s)= n\frac{\mathbb{E }|\Svec|^2 +\mathbb{E} |\vec{r}|^2 -\mathbb{E} |\vec{r}_{th}|^2}{R_s}$ unit dc sources. Dividing by the number of sources $n$, the unit dc we can distill out via the rectifier operation and combination in series is then
$$\frac{\mathbb{E }|\Svec|^2 +\mathbb{E} |\vec{r}|^2 -\mathbb{E} |\vec{r}_{th}|^2}{R_s}$$ per source. For the example of a hot resistor we have $\Svec=0$ and $\mathbb{E }|\vec{r}|^2=\mathbb{E }|\vec{r}_{T_H}|^2$, and the number of unit dc sources we can distill out should be $\frac{ \mathbb{E }|\vec{r}_{T_H}|^2-\mathbb{E }|\vec{r}_{th}|^2}{R_s}$.


\subsection{Formation of noisy sources from unit dc sources}
In the reverse direction, suppose we want to transform $m$ unit dc sources to e.g.\ $n$ hot resistors with $T>T_{ambient}:=T_{amb}$, as a  useful case. This can in principle be achieved by using the unit dc sources to amplify the thermal fluctuations of (free) thermal resistors with the same resistance as the target source and at temperature $T_{amb}$. Thermal noise is in fact normally amplified in order to be detected~\cite{Johnson1928}, so amplifying Gaussian noise is feasible in practice. For this combination of the unit dc sources and the thermal resistors to be a free operation, it is natural to require that it cannot increase the power that can be delivered to a load. $m$ unit dc sources can deliver $m$ joules into a resistive load at most (for a matched load). The original $n$ thermal resistors on their own deliver $$P_o=n\frac{\mathbb{E} |\vec{r}_{th}|^2}{R_s},$$
power and the $n$ target sources should deliver
$$P_T=n\frac{\mathbb{E}|\vec{ r}_{T}|^2}{R_s}$$
power. Thus the missing power, which naively equates to the required $m$ unit dc sources, is $P_m=n\frac{\mathbb{E}|\vec{ r}_T|^2-\mathbb{E }|\vec{r}_{th}|^2}{R_s}$, and 
$$\frac{m}{n}=\frac{\mathbb{E}|\vec{ r}_T|^2-\mathbb{E}|\vec{ r}_{th}|^2}{R_s}.$$
Thus the number of unit dc sources needed matches the value of the monotone $\unitdc$ of the target source as desired.


\section{Modelling noisy voltage sources given experimental data}
\label{sec:model}
One method of distilling out the signal from the noise is the averaging method, which works for AWGN by the law-of-large-number arguments used in the distillation section. The averaging function is built into many oscilloscopes \cite{horowitz1989art}.

In general one can use the maximum-likelihood approach. We now show how this can be used to justify a simple formula for determining the standard deviation of the AWGN noise from the observed data.  Let us consider that we are given with a data, say some microvoltages $\vec{x}=\{x_i\}_{i=1}^N$. The aim is to single out a signal  $\vec{S}$ and noise $\vec{r}$ from this data. To do so, we will take the approach of maximum-likelihood estimation \cite{Thomas2006}. Let us assume that the noise is modeled as additive white Gaussian noise with mean zero and standard deviation $\sigma$. Further, let us say that the actual signal is $\vec{S}=\{S_i\}_{i=1}^n$.
\begin{prop}
For AWGN with mean zero and standard deviation $\sigma$, the maximum-likelihood estimation of $\sigma$ for the proposed signal $\vec{S}$ and data $\vec{x}$ is given by
\begin{align}
\sigma=\sqrt{\frac{\vec{r}\cdot\vec{r}}{N}},
\end{align}
where $\vec{r}=\vec{x}-\vec{S}$ and $\vec{r}=\left(r_1,\cdots,r_N\right)^T$.
\end{prop}
\begin{mproof}
By assumption, the data $\vec{x}=\vec{S}+\vec{r}$. Then, assuming the noise to be an additive white Gaussian noise, the probability distribution $P(\vec{x},\sigma)$ is given by
\begin{align}
P(\vec{x},\sigma)=\prod_{i=1}^N\frac{1}{\sqrt{2\pi \sigma^2}} \exp\left[\frac{-(x_i-S_i)^2}{2\sigma^2}\right].
\end{align}
The maximum-likelihood approach selects the maximally likely distribution, which can be found by setting $\frac{\partial P(\vec{x},\sigma)}{\partial\sigma}=0$, which implies
\begin{align}
\sigma=\sqrt{\frac{\sum_{i=1}^N(x_i-S_i)^2}{N}}= \sqrt{\frac{\vec{r}.\vec{r}}{N}}.
\end{align}
This completes the proof of the proposition.
\end{mproof}



\end{document}